\documentclass[pre,aps,preprint,showpacs]{revtex4-1}

\usepackage{graphicx}
\usepackage{soul}
\newcommand{\dd}{{\protect\rm d}}

\usepackage{graphics} 
\usepackage{amssymb,amsmath}
\usepackage{amsmath}
\usepackage{psfrag}
\usepackage{graphicx}
\usepackage[]{xcolor}

\renewcommand{\b}{\beta}

\def\be{\begin{equation}} 
\def\ee{\end{equation}} 
\def\bea{\begin{eqnarray}} 
\def\eea{\end{eqnarray}}  
\def\bean{\begin{eqnarray*}} 
\def\eean{\end{eqnarray*}} 

\def\bse{\begin{subequations}}
\def\ese{\end{subequations}}

\def\lsim{\raise 0.4ex\hbox{$<$}\kern -0.8em\lower 0.62ex\hbox{$\sim$}} 
\def\gsim{\raise 0.4ex\hbox{$>$}\kern -0.7em\lower 0.62ex\hbox{$\sim$}}

\def\f0N{f_0^{(N)}}
\def\bec{\begin{center}}
\def\eec{\end{center}}

\begin{document}

\title{Classical Goldstone modes in Long-Range Interacting Systems}

\author{T.~M.~Rocha Filho}
\email{marciano@fis.unb.br}
\affiliation{Instituto de F\'{i}sica and International Center for Condensed Matter Physics,
Universidade de Bras\'\i lia, Campus Universit\'ario Darcy Ribeiro, Asa Norte, 70919-970 - Bras\'\i lia, Brazil}
\author{B.~Marcos}
\affiliation{Universit\'e C\^ote d’Azur, CNRS, Laboratoire J.-A.\ Dieudonn\'e, 06109 - Nice, France.}

\begin{abstract}
For a classical system with long-range interactions, a soft mode exists whenever a stationary state spontaneously
breaks a continuous symmetry of the Hamiltonian. Besides that, if the corresponding coordinate associated to the
symmetry breaking is periodic, the same energy of the different stationary states and finite $N$
thermal fluctuations result in a superdiffusive motion
of the center of mass for total zero momentum, that tends to a normal diffusion for very long-times.
As examples of this, we provide a two-dimensional self-gravitating system,
a free electron laser and the Hamiltonian Mean-Field (HMF) model.
For the latter, a detailed theory for the motion of the center of mass is given.
We also discuss how the coupling of the soft mode to the mean-field motion of individual particles may lead to strong chaotic
behavior for a finite particle number, as illustrated by the HMF model.
\end{abstract}


\maketitle

\section{Introduction}

Most of the literature on classical statistical mechanics and thermodynamics deals with systems  with short-range interparticle interactions,
in the sense that the interaction energy at interfaces is negligible with respect to the energy of the bulk of the system.
This ensures that energy, as well as entropy, are additive and extensive, two fundamental properties for the theoretical framework
of equilibrium statistical mechanics and thermodynamics~\cite{gibbs,ruelle,callen}. Yet many real systems fall outside this scope,
such as self-gravitating systems, charged plasmas, wave-plasma interaction, dipolar systems and
two-dimensional turbulence~\cite{physrep,levipprep,proc1,proc2,proc3,booklri} where the interaction is long-range, i.~e.\ with
an interparticle potential $v(r)$ that decays at large distances as $1/r^\alpha$, with $\alpha<d$ and $d$ the spatial dimension. As a consequence,
the total energy is no longer additive, which can lead to some interesting phenomena as ensemble-inequivalence,
negative specific heat, non-Gaussian stationary states (in the limit of an infinite number of particles),
and more importantly for the present work, anomalous diffusion.

Let us consider an $N$-particle systems with Hamiltonian:
\begin{equation}
H=\sum_{i=1}^N\frac{{\bf p}_i^2}{2m}+\frac{1}{N}\sum_{i<j=1}^Nv(|{\bf r}_i-{\bf r}_j|),
\label{genham}
\end{equation}
where ${\bf r}_i$ and ${\bf p}_i$ are the position and conjugate momentum of the $i$-th particle, respectively.
The $1/N$ factor in the potential energy term is a Kac factor~\cite{kac} introduced for the energy to be extensive
quantity (on this point see for instance the discussion in chapter 2 of~\cite{booklri}).
Under suitable conditions, in the $N\rightarrow\infty$ limit the dynamics described by the Hamiltonian in Eq.~(\ref{genham})
is mathematically equivalent to a mean-field description  with the one-particle distribution function satisfying
the Vlasov equation~\cite{braun,steiner,brenig}, i.~e.\ all particles are uncorrelated.


If the original Hamiltonian is
invariant with respect to translation of one coordinate, and the equilibrium (or stationary) state spontaneously breaks
this symmetry, then a soft mode, i.~e.\ a Goldstone mode, exists with zero energy cost to go from one equilibrium state
to another~\cite{goldstone,goldstone2,martin}. Besides, if the coordinate associated to the broken symmetry is periodic, then
thermal excitations of this soft mode lead to a diffusion of the center of mass of the equilibrium state, as discussed below.
Our aim in the present work is then to show how classical Goldstone modes are realized in long-range interacting systems when
a symmetry of the Hamiltonian is broken, either for an equilibrium or a non-equilibrium stationary state, and how, in the case
of a cyclic coordinate, thermal fluctuations lead to a superdiffusive, ballistic in an initial regime, motion of the center of mass of the system.
This behavior is expected to be ubiquitous for all systems with long-range interactions and periodic coordinates, under the stated conditions.
We illustrate this phenomenology for three paradigmatic models with long-range interactions: the Hamiltonian Mean Field (HMF)
model~\cite{booklri,hmforig}, two-dimensional self-gravitating particles~\cite{miller} and the single pass free electron
laser~\cite{booklri,bonifacio,yves1,yves2,yves3}. Due to its inherent simplicity, yet retaining the main characteristics
of systems with long-range interactions, the HMF model has been extensively studied in the literature. This simplicity
will allow us here to present a more detailed theoretical description of this soft mode and of the
superdiffusive motion of the center of mass of the system.

The paper is structured as follows: In Section~\ref{sec2} we explain the physical mechanism for the diffusive motion of the center of mass
of a statistical stationary state, the thermal excitation of the Goldstone mode, and its relation to the diffusion of individual particles.
In Section~\ref{sec3} we illustrate this for the HMF model, for both equilibrium and non-equilibrium states, and present a theoretical approach
for determining the properties of the diffusive motion of the center of mass.
The enhancement of chaos due to the presence of the soft mode
is discussed in Sec.~\ref{sec6} and illustrated for the HMF model.
In Section~\ref{secother} we shows that the same diffusive motion of the center of mass is observed in two other
systems with long-range interactions:  a two-dimensional self-gravitating system and a free electron laser,
illustrating the generality of this behavior.
We close the paper with some concluding remarks and perspectives in Sec.~\ref{sec7}.

\section{Goldstone Modes in Classical Statistical Mechanics of Systems with Long-Range Interactions}
\label{sec2}

Spontaneous symmetry breaking is one of the landmarks of the developments of theoretical physics in the last
half-century, occurring from subatomic up to macroscopic systems~\cite{goldstone,goldstone2},
as exemplified by the Brout-Englert-Higgs phenomenon, superconductivity, soft-mode turbulence, phonons in solids,
and plasmons, among others~\cite{goldstone,goldstone2,morchio,rossberg}.
Although usually first introduced for quantum systems, Goldstone modes can also be defined in a classical
context~\cite{strocchi1,strocchi2}, provided a few conditions are met. The system must have an infinite
number of degrees of freedom, with its dynamics having the property that the space of physical states is
divided in disconnected islands stable under time evolution. Here disconnected means that a state
from one island cannot be reached from a state of a different island by physically realizable process
without external intervention.
In Statistical Mechanics, each island corresponds to a given state of thermodynamic equilibrium
(which is not unique for a given energy if a symmetry is broken),
and all those states that evolve into it.
For long-range interacting systems one has to also consider islands associated
to stationary states other than the Maxwell-Boltzmann (MB) equilibrium distributions.
Indeed, in the thermodynamics limit, there are an infinite number of such non-Gaussian states which
never evolve to equilibrium, and as a consequence, each such state is part of a disconnected island, again with all states
that evolve towards it, in the same sense as for equilibrium states. A symmetry breaking
occurs in a given island when it is unstable by  the operation of a symmetry subgroup of the whole
symmetry group of the system (the symmetries of the Hamiltonian). The Goldstone Theorem for classical
systems then states (see Ref.~\cite{strocchi2} for additional mathematical details) that, for each
broken symmetry in a given island, there exists a solution of the dynamics satisfying the free wave
equation (Goldstone modes).

For a finite but still a large number of particles $N$, the islands referred above are no longer,
strictly speaking, invariant under the system dynamics. A stationary state for finite $N$ acquires
a life-time and is now called a Quasi-Stationary State (QSS) and can leave an island by evolving in time
into the final MB thermodynamic equilibrium~\cite{booklri,scaling,scaling2}.
Although the invariance of the islands is lost, the time scale, i.~e.\ the relaxation time over which
the QSS evolves is typically very large, and one can still consider the free wave solution states
as long lived Goldstone modes, that slowly relax to the mode corresponding to the final equilibrium state,
as discussed below.

\begin{figure}[ptb]
\begin{center}
\scalebox{0.4}{{\includegraphics{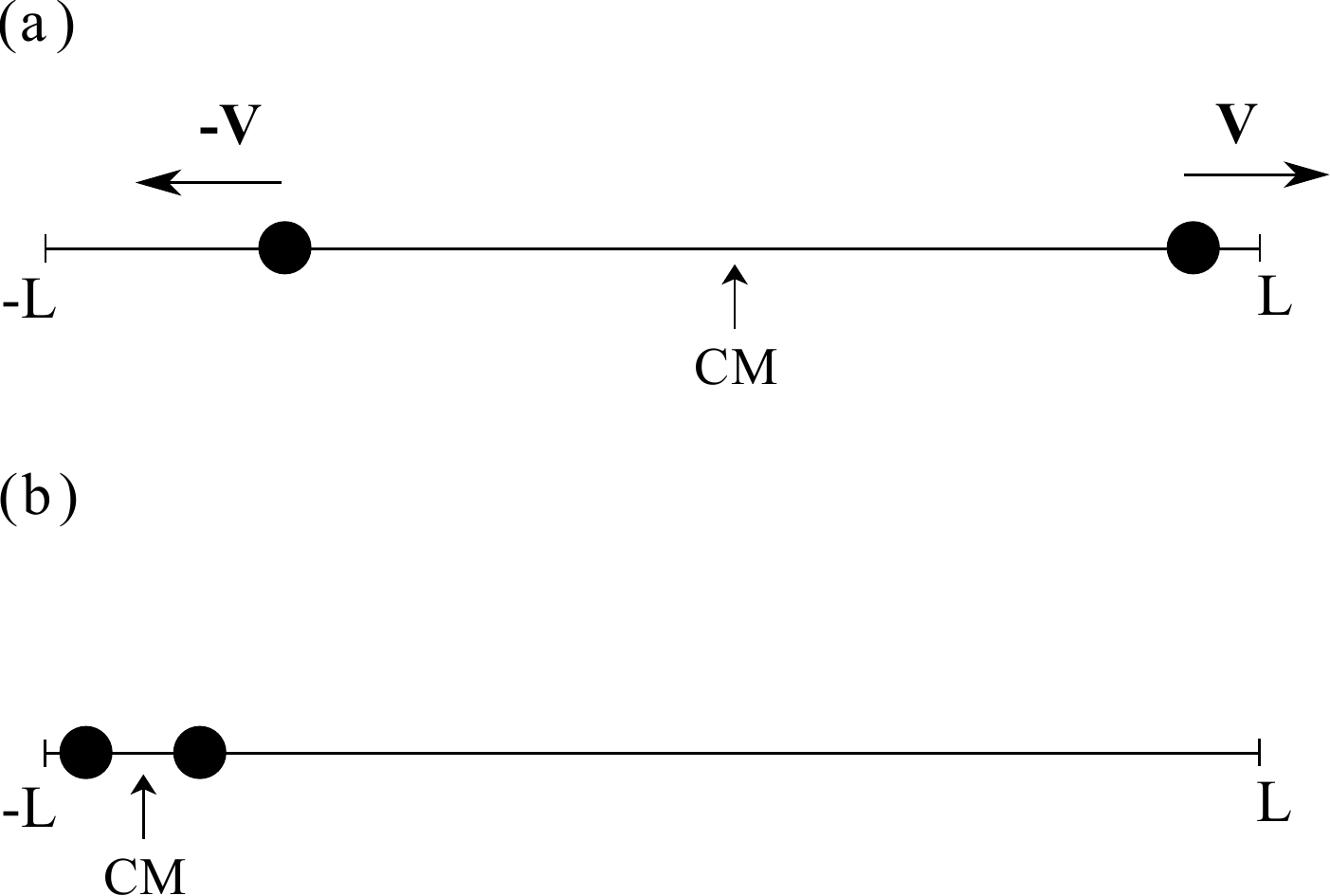}}}
\end{center}
\caption{We consider two particles with a periodic coordinate in the interval $[-L,L)$:
a) In the initial state the total momentum vanishes as both particles have opposite velocities, and the position
of the center of mass (CM) is indicated by the vertical arrow. b) The particles have moved freely,
but one of the particles reaches one boundary before the other, and appears at the other side of the periodic
one-dimensional space. As a consequence, the center of mass is now at a different position.}
\label{ex_periodic}
\end{figure}

Here we are interested in Goldstone modes realized in long-range systems with periodic boundary condition
(described using a periodic coordinate). For that purpose, let us suppose that the energy is invariant
under translations of a periodic coordinate
$\theta$ with periodicity $2\pi$, with conjugate momentum $p_\theta$, and that
the system is in a (quasi-) stationary state or in the true thermodynamic equilibrium.
If such a state spontaneously breaks the translation symmetry with respect to $\theta$
for a finite number of particles $N$, then the corresponding Goldstone
and thermal fluctuations due to the finite number of particles results in a diffusive motion of
the center of mass of the system with vanishing total momentum (see below).
This is not a contradictory statement as illustrated by the simple example in Fig.~\ref{ex_periodic}.
We observe that this is a completely different phenomenon from the non-conservation of angular momentum in simulations
with artificial periodic boundary conditions~\cite{kuzkin}. In the latter case, periodicity is a non-physical
computational artifact to simplify numerical simulations, and has as a side-effect the non-conservation
of angular momentum. Here angular momentum is always strictly conserved and the periodic boundary is truly physical.

The equilibrium state (or a quasi-stationary state) with zero average momentum is represented by the distribution function $f_0(\theta,p)$,
considered to be centered initially at $\theta=0$, with fluctuations described by $\delta f(\theta,p;t)$,
that can be considered to be of order $1/\sqrt{N}$ and preserving the total (zero) momentum, i.~e.
\begin{equation}
\int_{-\pi}^\pi\dd \theta\int_{-\infty}^\infty\dd p f_0(\theta,p)=1,
\label{flucmompr1}
\end{equation}
\begin{equation}
\int_{-\pi}^\pi\dd \theta\int_{-\infty}^\infty\dd p\:p\: f_0(\theta,p)=0,
\label{zeromom}
\end{equation}
and
\begin{equation}
\int_{-\pi}^\pi\dd \theta\int_{-\infty}^\infty\dd p\: \delta f(\theta,p;t)=\int_{-\pi}^\pi\dd \theta\int_{-\infty}^\infty
\dd p\:p\: \delta f(\theta,p;t)=0,
\label{flucmompr}
\end{equation}
with $f_0+\delta f\geq0$.
As $\theta\in[-\pi,\pi)$ with periodic boundary conditions, we denote the number of particles per unit of time
crossing from positive values of $\theta$ at the boundary at $\theta=\pi$ as $N_+$ and the particles crossing by unit of time
from negative values of $\theta$ at $\theta=-\pi$ as $N_-$. We then have that:
\begin{equation}
N_+=\int_ 0^\infty\dd p\left[f_0(\pi,p)+\delta f(\pi,p,t)\right]p,
\label{NplusL}
\end{equation}
and
\begin{equation}
N_-=-\int_ {-\infty}^0\dd p\left[f_0(-\pi,p)+\delta f(-\pi,p,t)\right]p.
\label{NminusL}
\end{equation}
The net flux of particles at the boundary $\theta=\pi$ is then given by
\begin{equation}
\Delta N=N_+-N_-=\int_0^\infty\dd p\left[\delta f(\pi,p;t)-\delta f(-\pi,-p;t)\right]p+\int_{-\infty}^\infty\dd p f_0(\pi,p)p,
\label{netflux1}
\end{equation}
where we used explicitly the periodicity in space of $f_0(\theta,p)$. The last term in the right-hand
side of Eq.~(\ref{netflux1}) vanishes identically, which is equivalent to say that the net flux of particles at the borders for
the unperturbed distribution $f_0$ is zero. Using the fact that $\delta f$ must also be periodic in $\theta$, we obtain:
\begin{equation}
\Delta N=\int_0^\infty\dd p\left[\delta f(\pi,p;t)-\delta f(\pi,-p;t)\right]p.
\label{netflux2}
\end{equation}
The important point is that $\delta f(\pi,p;t)$ does not have to be equal to $\delta f(\pi,-p;t)$,
but yet complying with a total vanishing momentum.
This shows that the periodic boundary conditions together with a non-symmetric fluctuation with respect to $p$ implies a net movement of the
stationary state, which is governed by the nature of finite $N$ fluctuations.

The time derivative of the position of the center of mass $\phi\equiv\langle\theta\rangle$ is then obtained from the considerations
in the previous paragraph as:
\begin{equation}
\dot\phi(t)=-\frac{2\pi}{N}\Delta N=-\frac{2\pi}{N}\int_0^\infty\dd p\left[\delta f(\pi,p;t)-\delta f(\pi,-p;t)\right]p.
\label{derxcm}
\end{equation}
To show that the motion of the center of mass corresponds to a diffusive process, we write the variance of its position as
\begin{equation}
\sigma_\phi^2(t)=\left\langle\left[\phi(t)-\phi(0)\right]^2\right\rangle,
\label{cmvar}
\end{equation}
where
\begin{equation}
\phi(t)=\frac{1}{N}\sum_{i=1}^N\theta_i(t)
\label{cmposdef}
\end{equation}
and
 $\langle\cdots\rangle$ stands for an average over different realizations for the same (macroscopic) initial state.
By choosing the origin such that $\phi(0)=0$ we have
\begin{equation}
\sigma_\phi^2(t)=\left\langle\left[\frac{1}{N}\sum_{i=1}^N\theta_i(t)\right]^2\right\rangle
=\frac{1}{N^2}\left\langle\sum_{i=1}^N\theta_i^2(t)\right\rangle+\frac{1}{N^2}
\left\langle\sum_{\substack{i,j=1\\ i\neq j}}^N\theta_i(t)\theta_j(t)\right\rangle.
\label{difcmdiffpart}
\end{equation}
Although the position angles are restricted to the interval $[-\pi,\pi)$, for considering diffusive processes it is useful to
consider both the center of mass and particle position to evolve on the whole real axis, and from now, we define $\phi$ in this way.
By folding back to the original interval we recover the motion on the circle.
We now note that interparticle correlations for a long-range interacting system with a potential regularized by a Kac factor
are of order $1/N$~\cite{steiner}, and therefore $\langle\theta_i\theta_j\rangle=\langle\theta_i\rangle\langle\theta_j\rangle+{\cal O}(1/N)$.
Since the average of the position of any particle over many realization must vanish by construction, the last term in the right-hand side
of Eq.~(\ref{difcmdiffpart}) is of order $1/N^3$ and is therefore negligible for large $N$.
From the definition of the variance of the position of the particles in the system:
\begin{equation}
\left\langle\frac{1}{N}\sum_{i=1}^N\theta_i^2\right\rangle=\sigma_\theta^2,
\label{partvar}
\end{equation}
we thus have that:
\begin{equation}
\sigma_\phi^2(t)=\frac{1}{N}\sigma_\theta^2(t).
\label{difcmdiffpart2}
\end{equation}
The particles are initially confined in the interval $-\pi\leq\theta<\pi$, and since typically $|\theta|$ gets much greater
than $\pi$ with time, we can write with a minor error that becomes negligible with increasing time that
\begin{equation}
\sigma_\theta^2\rightarrow\frac{1}{N}\sum_{i=1}^N\left[\theta_i(t)-\theta_i(0)\right]^2.
\label{aproxdiff}
\end{equation}
We conclude that the diffusion of center of mass of the system is due to the diffusion of individual
particles viewed as interacting on an infinite space with a periodic interparticle potential. As a consequence,
the dynamics of center of mass position can be described by the same type of equations that describe the diffusion in
the system. For instance, if a Langevin equation is known for the motion of a single particle, then a corresponding
Langevin equation can be written for the center of mass by a simple rescaling by a factor $1/N$.
The study of diffusion in position for particles with long-range interactions is not a simple task and was studied
in the literature, but a more complete theory is still lacking (see~\cite{yamaguchi2,difus1,difus2,difus2b,difus3,difus4} and references therein).
However, for the much studied HMF model, a more detailed description of the phenomenon is possible for the
initial ballistic diffusion regime, as will be shown in the next section.

\section{The Hamiltonian Mean Field Model}
\label{sec3}

The HMF model is formed by $N$ particles on a ring globally coupled by a cosine potential and Hamiltonian~\cite{booklri,hmforig}:
\begin{equation}
H=\sum_{i=1}^N \frac{p_i^2}{2}+\frac{1}{N}\sum_{i<j=1}^N \left[1-\cos(\theta_i-\theta_j)\right].
\label{hmfham}
\end{equation}
This model is widely studied in the literature due to its inherent simplicity. Particularly, due to the form of its interparticle potential
the numerical effort in molecular dynamics simulations scales linearly with $N$, instead of $N^2$, which allows very long simulation times
for very large number of particles (see Refs.~\cite{physrep,eu2} and references therein).
The magnetization components for the HMF model are defined by:
\begin{equation}
M_x=\frac{1}{N}\sum_{i=1}^N \cos(\theta_i),\hspace{5mm}M_y=\frac{1}{N}\sum_{i=1}^N \sin(\theta_i),
\label{hmfmags}
\end{equation}
and the total magnetization by $M=\sqrt{M_x^2+M_ y^2}$. The system is solvable and the
one particle equilibrium distribution is given by~\cite{hmforig,eu3}:
\begin{equation}
f_{\rm eq}(\theta,p)=\frac{\sqrt{\beta}}{(2\pi)^{3/2}I_0(\beta M)}
\exp{\left\{-\beta\left[\frac{p^2}{2}-M_x\cos(\theta)-M_y\sin(\theta)\right]\right\}},
\label{hmfeqstate}
\end{equation}
where $I_k$ is the modified Bessel function of the first kind with index $k$.
The magnetization $M$ as a function of the inverse temperature $\beta$ is obtained from the solution of the equation:
\begin{equation}
M=\frac{I_1(\beta M)}{I_0(\beta M)}.
\label{hmfeqeq}
\end{equation}
We denote the total energy per particle as $e\equiv H/N$, with $H$ the total Hamiltonian of the system.
The system has a second order phase transition from a ferromagnetic phase at lower energies to a homogeneous non-magnetic phase
at higher energies with a critical energy per particle $e=0.75$.
Since only the modulus $M$ is determined for a given temperature,
the equilibrium state is infinitely degenerate for $M\neq0$, and the
rotational symmetry of the total Hamiltonian is spontaneously broken.

\begin{figure}[ptb]
\begin{center}
\scalebox{0.3}{{\includegraphics{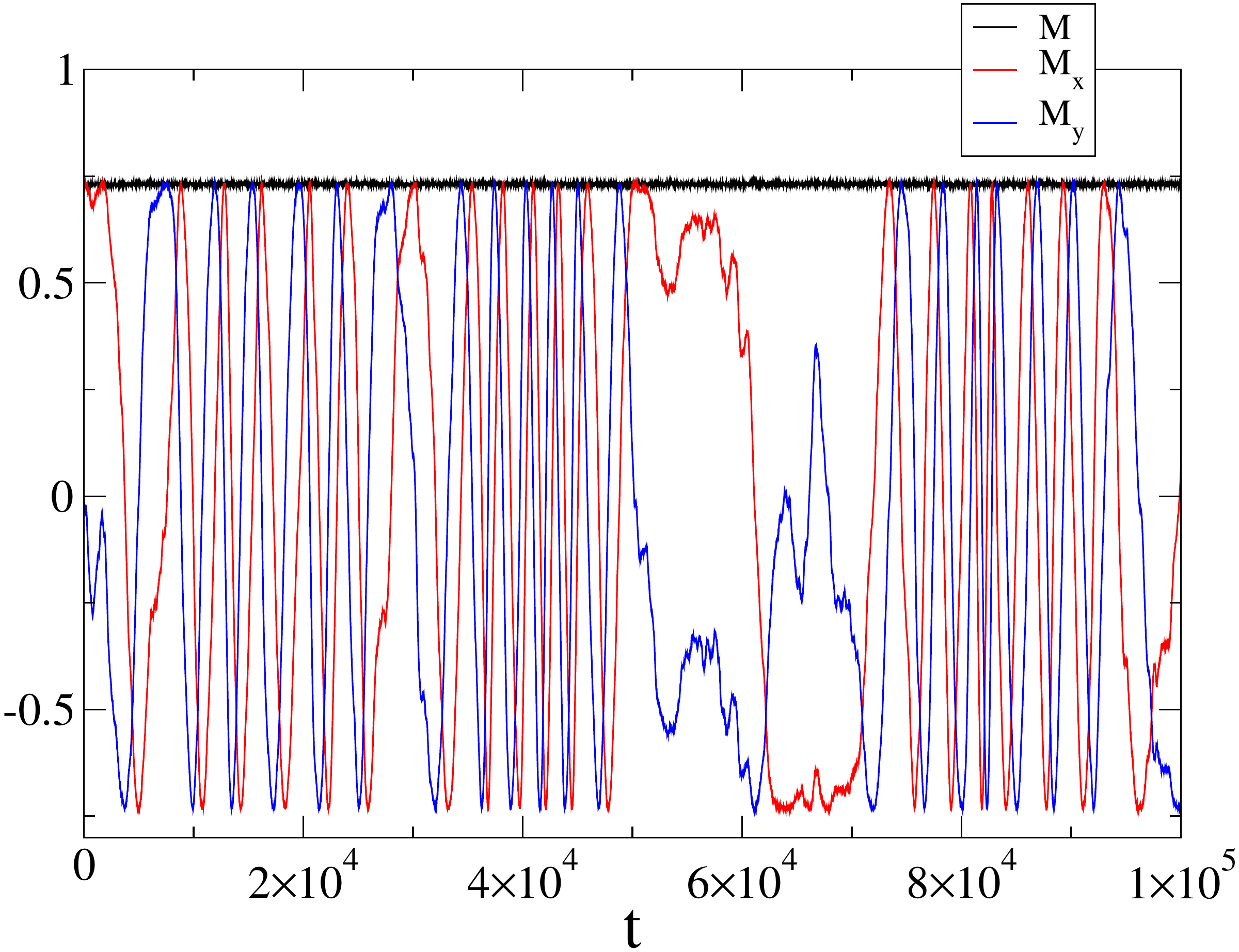}}}
\scalebox{0.3}{{\includegraphics{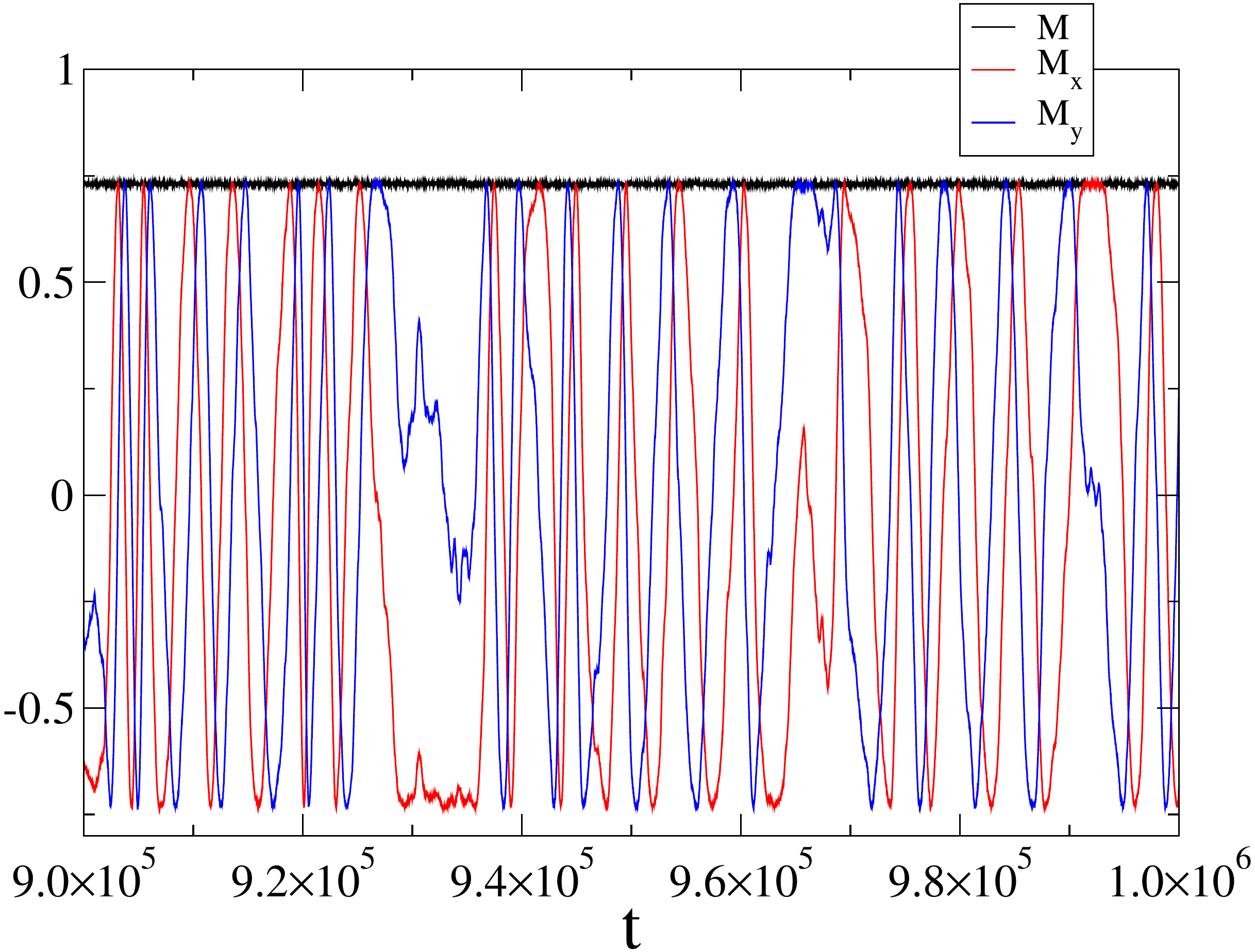}}}
\end{center}
\caption{(Color online) Total magnetization and component $M_x$ and $M_y$ for the
HMF model at thermodynamic equilibrium for two time windows,
with energy per particle $e=0.4$, $N=10\,000$ particles, time step $\Delta t=0.5$ and energy relative error
of order $10^{-4}$.}
\label{hmf_mag_10000}
\end{figure}

\begin{figure}[ptb]
\begin{center}
\scalebox{0.3}{{\includegraphics{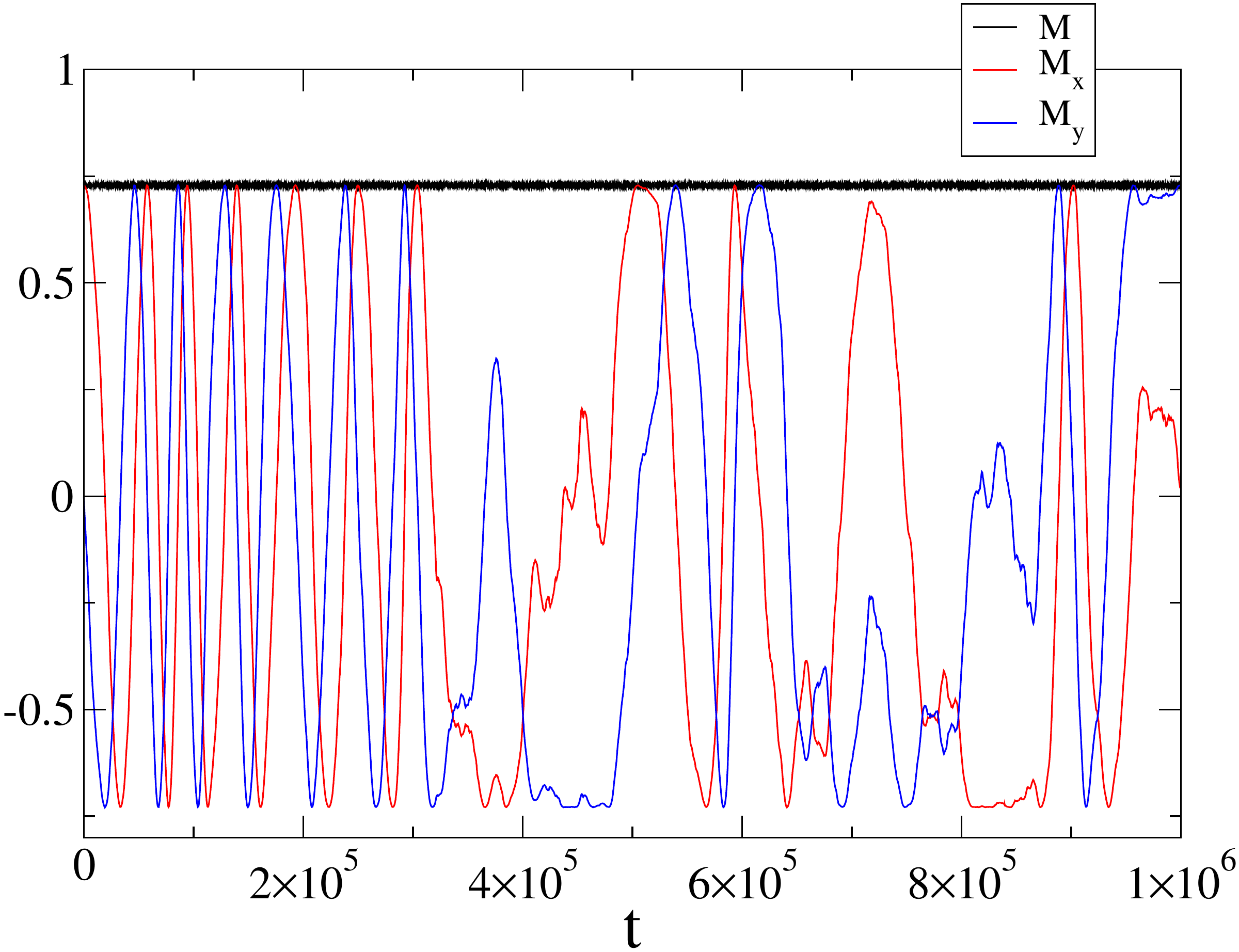}}}
\scalebox{0.3}{{\includegraphics{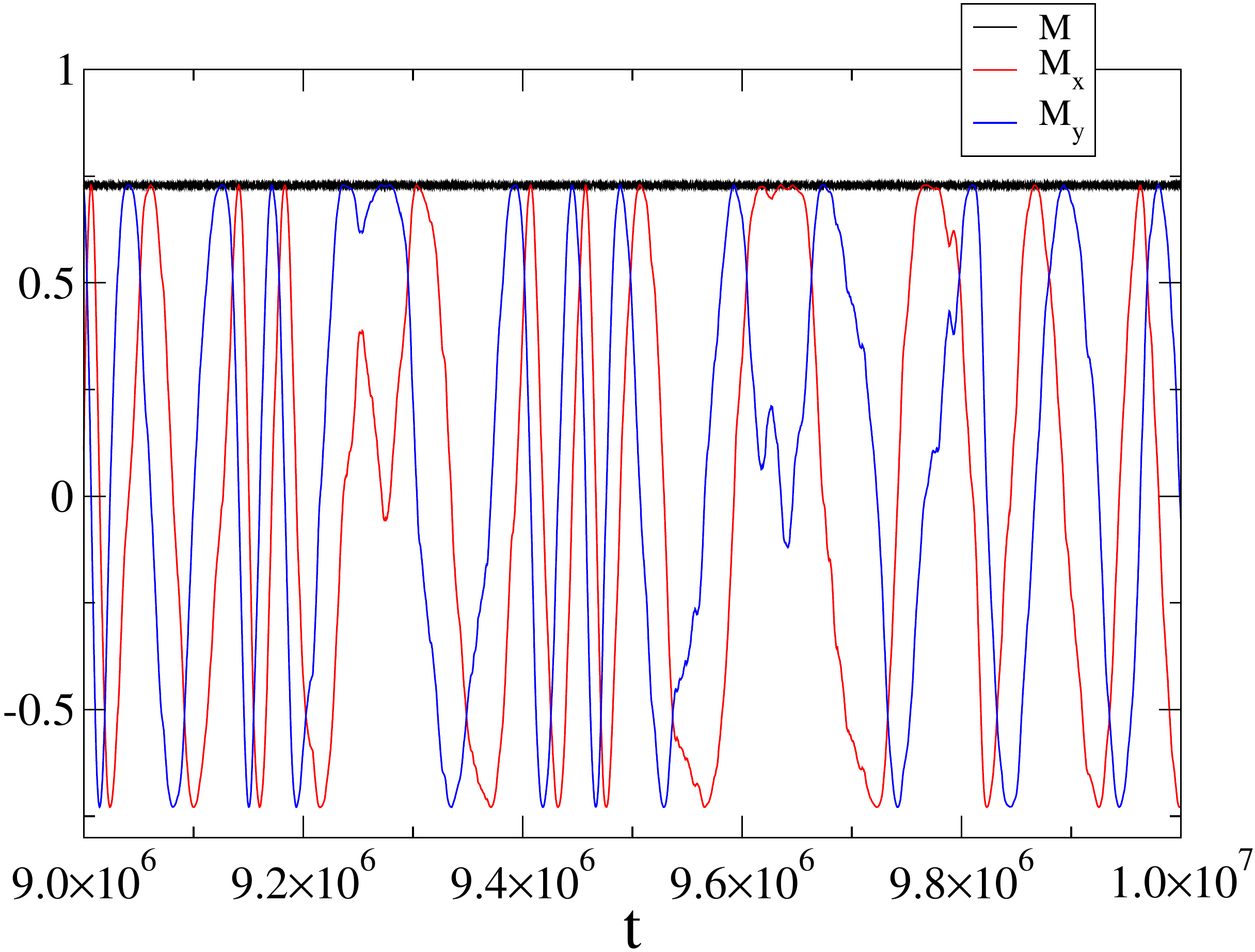}}}
\end{center}
\caption{(Color online) Same as Fig.~\ref{hmf_mag_10000} but with $N=1\,000\,000$
particles and energy relative error of order $10^{-5}$
and final total momentum per particle of order $10^{-7}$.}
\label{hmf_mag_1000000}
\end{figure}

\begin{figure}[ptb]
\begin{center}
\scalebox{0.3}{{\includegraphics{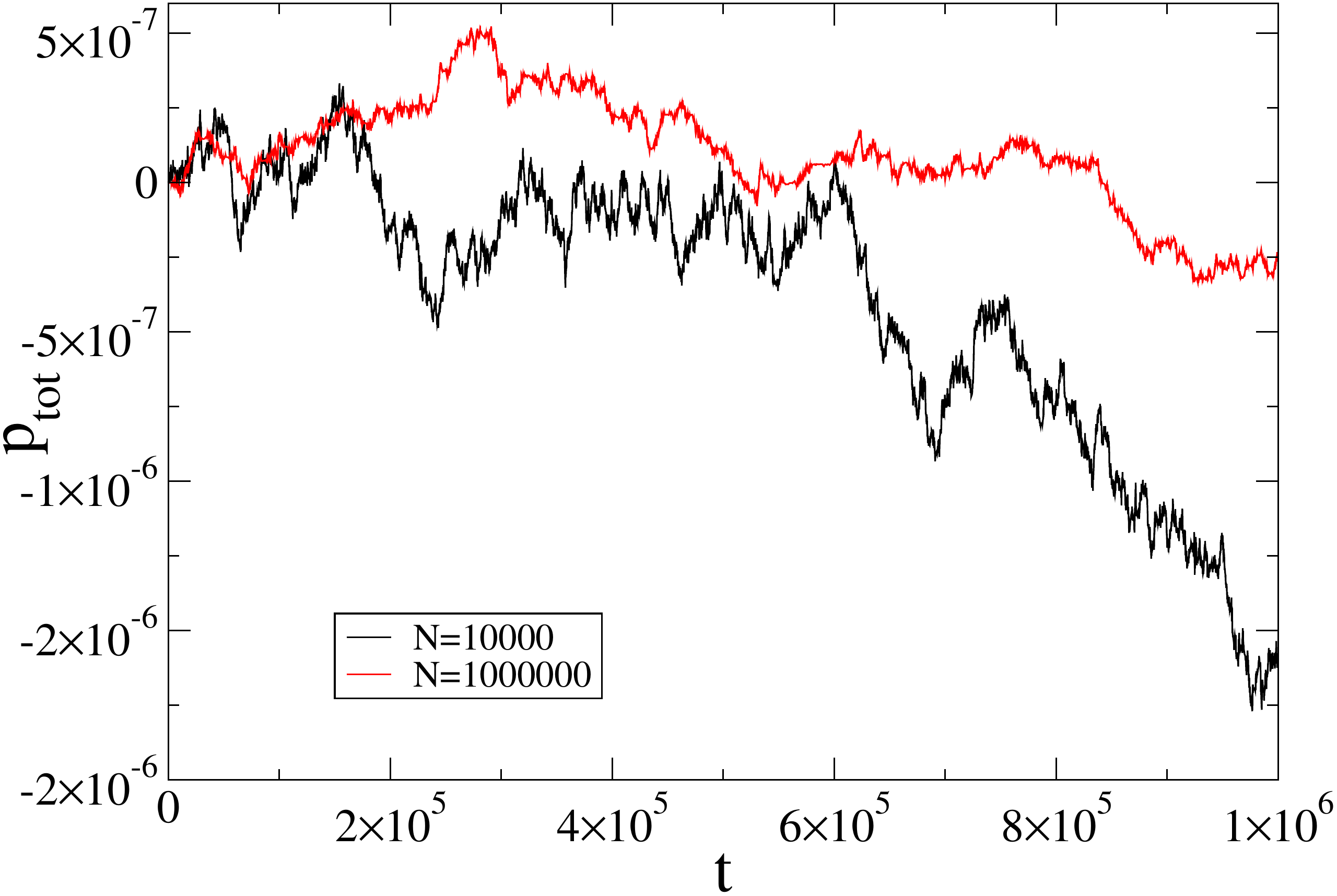}}}
\end{center}
\caption{(Color online) Total momentum per particle for the simulations in Figs.~\ref{hmf_mag_10000} and~\ref{hmf_mag_1000000}.}
\label{hmdPtoterr}
\end{figure}

\begin{figure}[ptb]
\begin{center}
\scalebox{0.3}{{\includegraphics{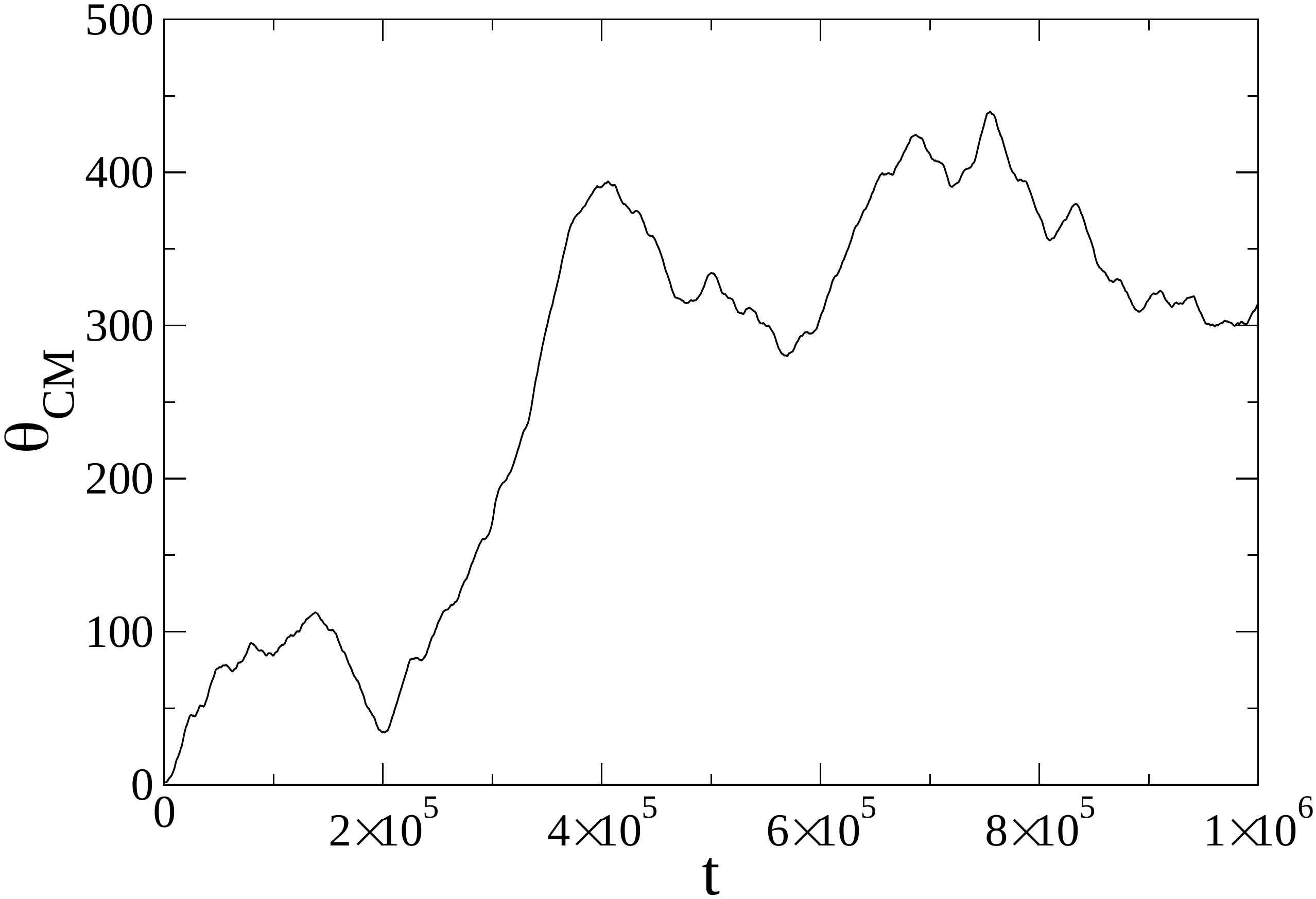}}}
\scalebox{0.3}{{\includegraphics{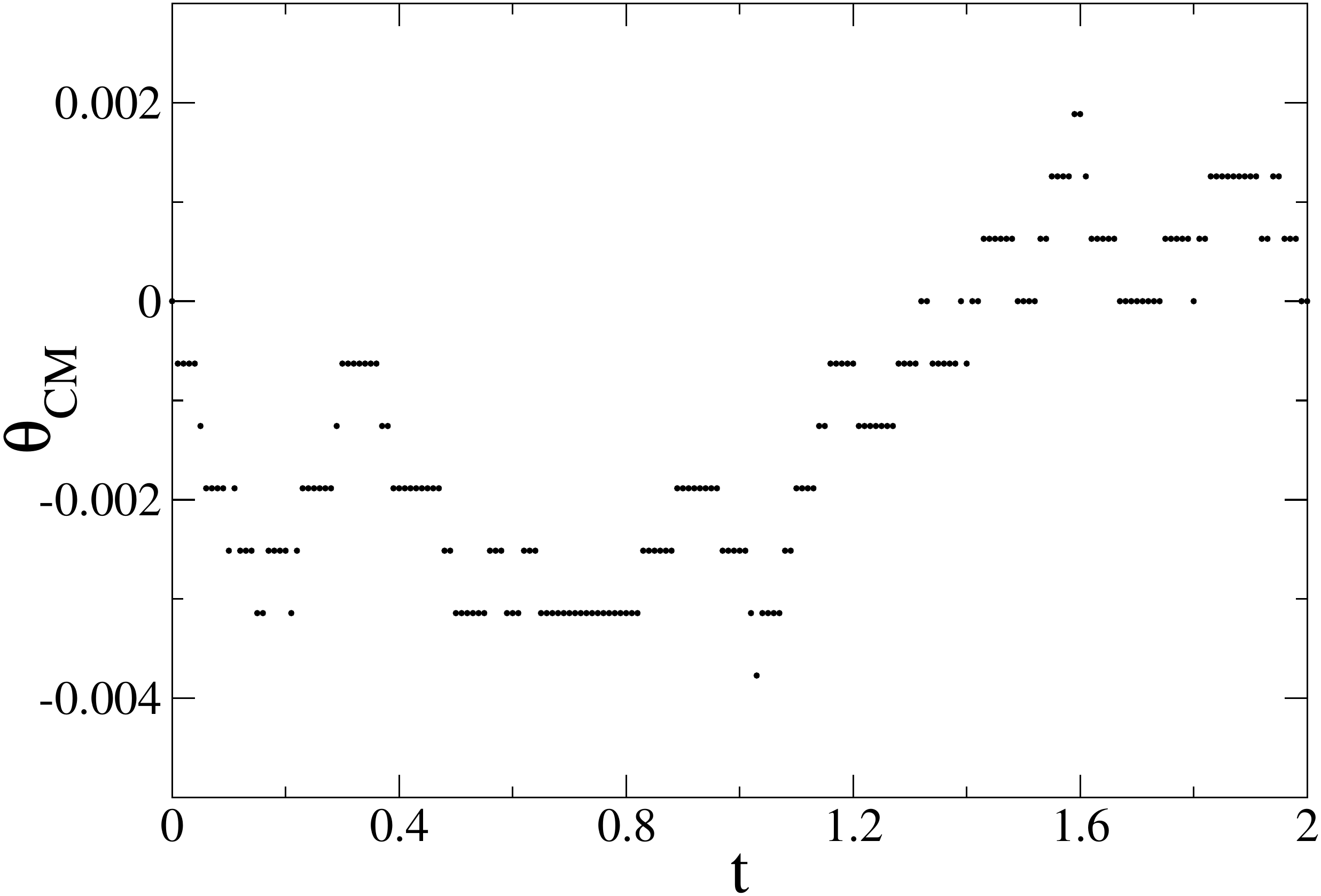}}}
\end{center}
\caption{Left Panel: Center of mass of particles for the same case as in Fig.~\ref{hmf_mag_10000} but with a time step of $\Delta t=10^{-2}$.
Right Panel: Zoom over the initial portion of the graphic in the left panel, showing the discrete nature of the center of mass motion.}
\label{hmf_alpha}
\end{figure}

As the thermodynamic limit is equivalent to the mean-field description and particles are uncorrelated~\cite{braun},
it is straightforward to show that the time derivatives of $M_x$ and $M_y$ vanish.
Nevertheless, for finite $N$, small correlations are present and result in a slow variation of the magnetization components with time.
Figures~\ref{hmf_mag_10000} and~\ref{hmf_mag_1000000} show the time evolution of the magnetization components, with a
total constant magnetization up to small fluctuations, for an equilibrium magnetized
(non-homogeneous) state for $N=10\,000$ and $1000\,000$, and total energy per particle $e=0.4$.
The total momentum remains zero and constant up to very small numeric errors as shown in Fig.~\ref{hmdPtoterr}.
Figure~\ref{hmf_alpha} shows the displacement of the angular position of the center of mass,
which coincides with the phase of the magnetization given by
$M_x + i M_y = M\exp(i\phi)$, for the case in Fig.~\ref{hmf_mag_10000} for $N=10000$,
with a typical diffusive random motion behavior. The discrete nature of this motion is evidenced on the right-panel of Fig.~\ref{hmf_alpha},
as the center of mass jumps by $\pm\pi/N$ for each particle traversing the periodic boundary.
Comparing Figs.~\ref{hmdPtoterr} and~\ref{hmf_alpha} it is evident that the motion of the
center of mass is orders of magnitude bigger that would be expected from the small errors in the numeric integrator.
The oscillations are quasi-periodic with chaotic intermittencies and never damp, as the long time window of the simulation shows clearly.
For all times the system is in a degenerate equilibrium state, with a time varying position of its center of mass caused by thermal fluctuations
for finite $N$.
This time dependence of the phase of the magnetization was first noted for the HMF model by Ginelli et al.\ in Ref.~\cite{ginelli},
and also by Manos and Ruffo relating it to the transition from weak to strong chaos for the same model~\cite{manos}. We will discuss this
last point with more details in Section~\ref{sec6}.

\begin{figure}[ptb]
\begin{center}
\scalebox{0.3}{{\includegraphics{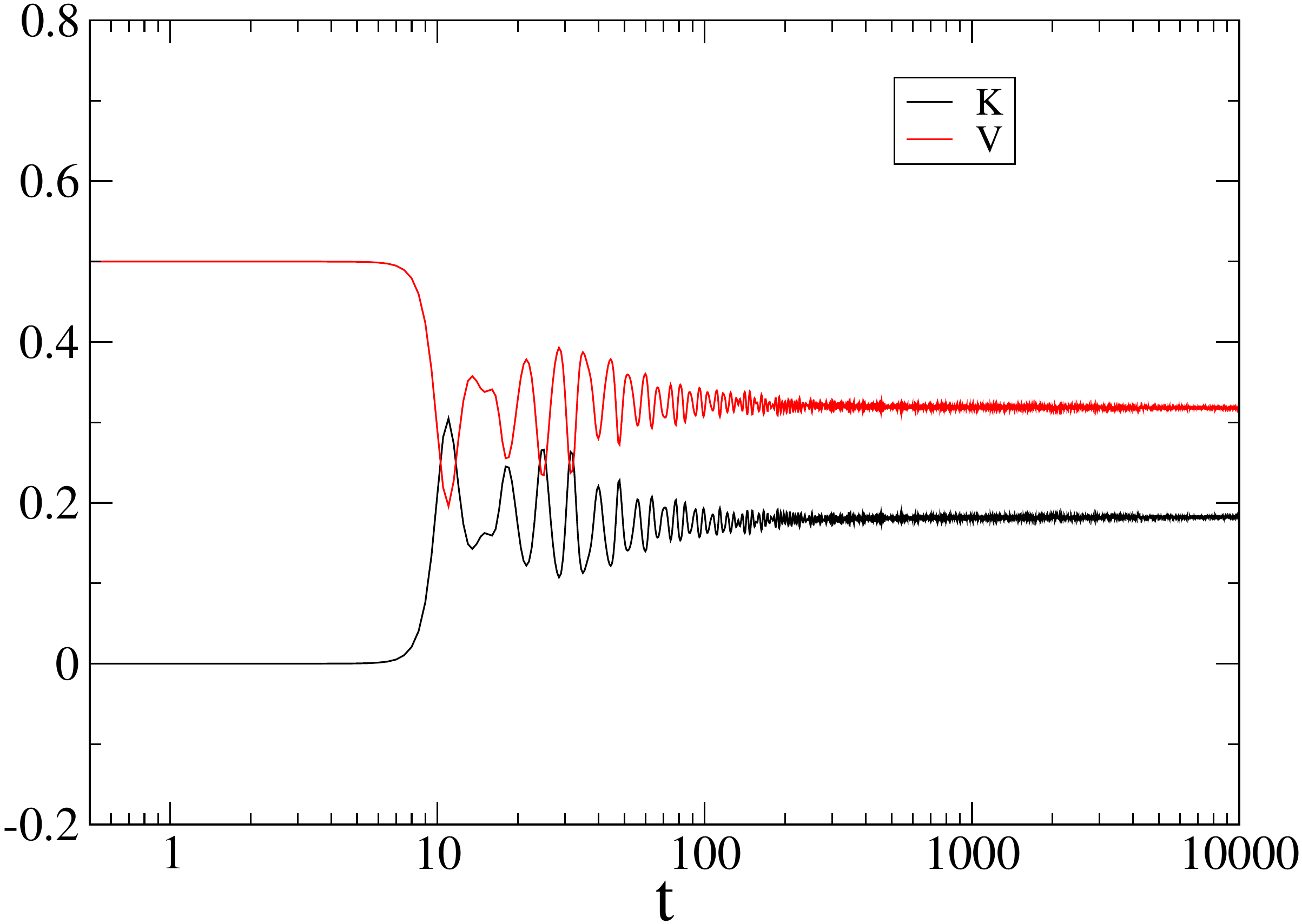}}}
\scalebox{0.3}{{\includegraphics{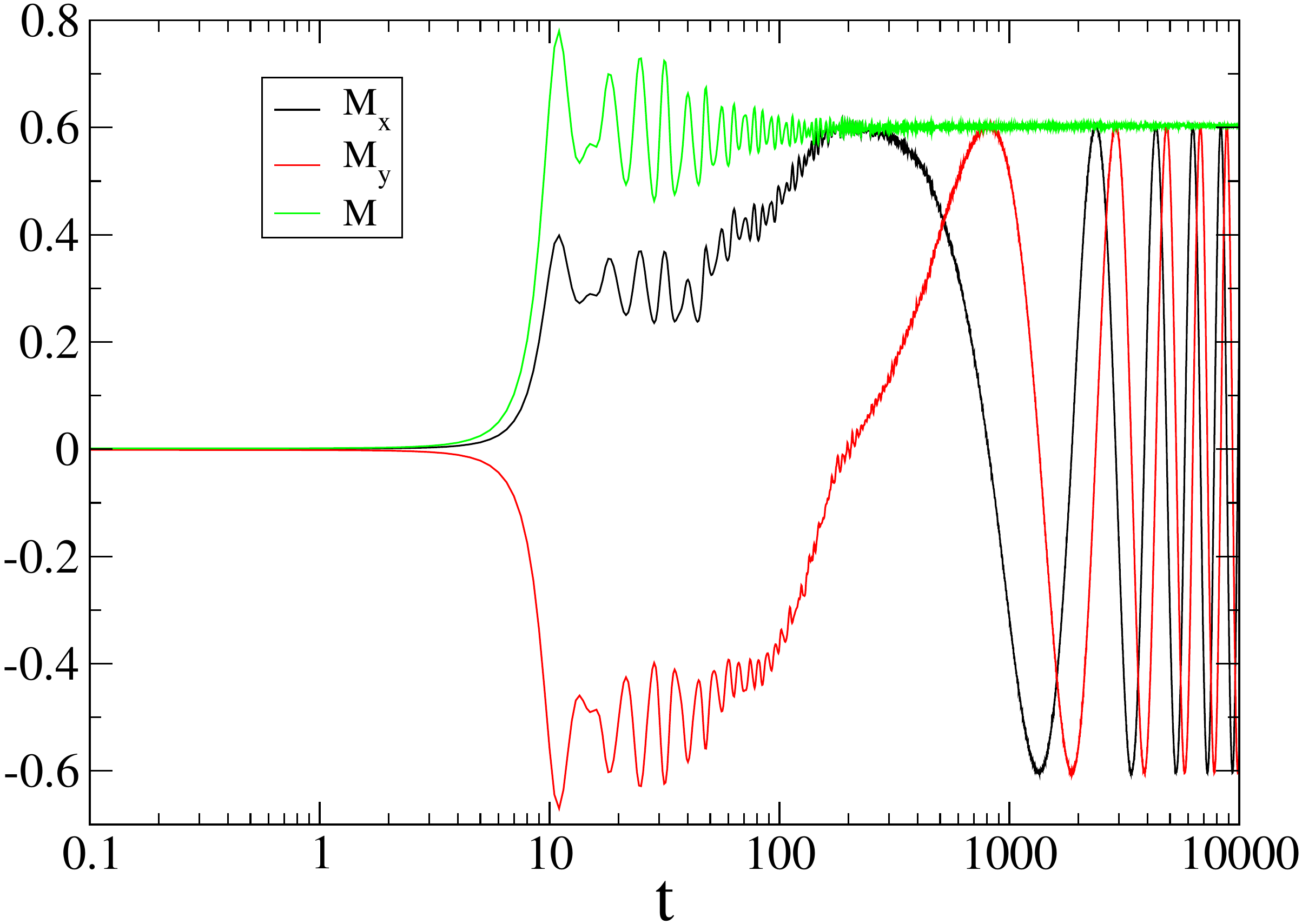}}}
\end{center}
\caption{(Color online) Left panel: Kinetic ($K$) and potential ($V$) energies
per particle for an out of equilibrium evolution of waterbag initial state with total energy per particle $e=0.5$ and initial magnetization
$M_0=0$ of the HMF model, with $N=1\,000\,000$. Right panel: total magnetization and its components corresponding to the left panel.
The initial violent relaxation is clearly visible, as well as the final oscillatory behavior of the magnetization.}
\label{HMFnoneq}
\end{figure}

Non-equilibrium states also display the same behavior for finite $N$ as long as the magnetization is not zero.
Let us take as initial condition a waterbag state:
\begin{equation}
	f(p,\theta)=
	\left\{
		\begin{array}{l}
			1/(4p_0\theta_0),\:\:{\rm if}\:\:-p_0<p<p_0\hspace{2mm}{\rm and}\hspace{2mm}-\theta_0 < \theta < \theta_0;
			\\
			0,\:\:{\rm otherwise}.
		\end{array}
		\right.
	\label{wbstate}
\end{equation}
Figure~\ref{HMFnoneq} shows the dynamical evolution of an initial unstable waterbag state with $M=0$ ($\theta_0 =\pi$). It goes though an initial
violent relaxation and then settles into a magnetized quasi-stationary state, with a time varying phase of the magnetization similar to the what
is observed at thermodynamic equilibrium.


\begin{figure}[ptb]
\begin{center}
\scalebox{0.3}{{\includegraphics{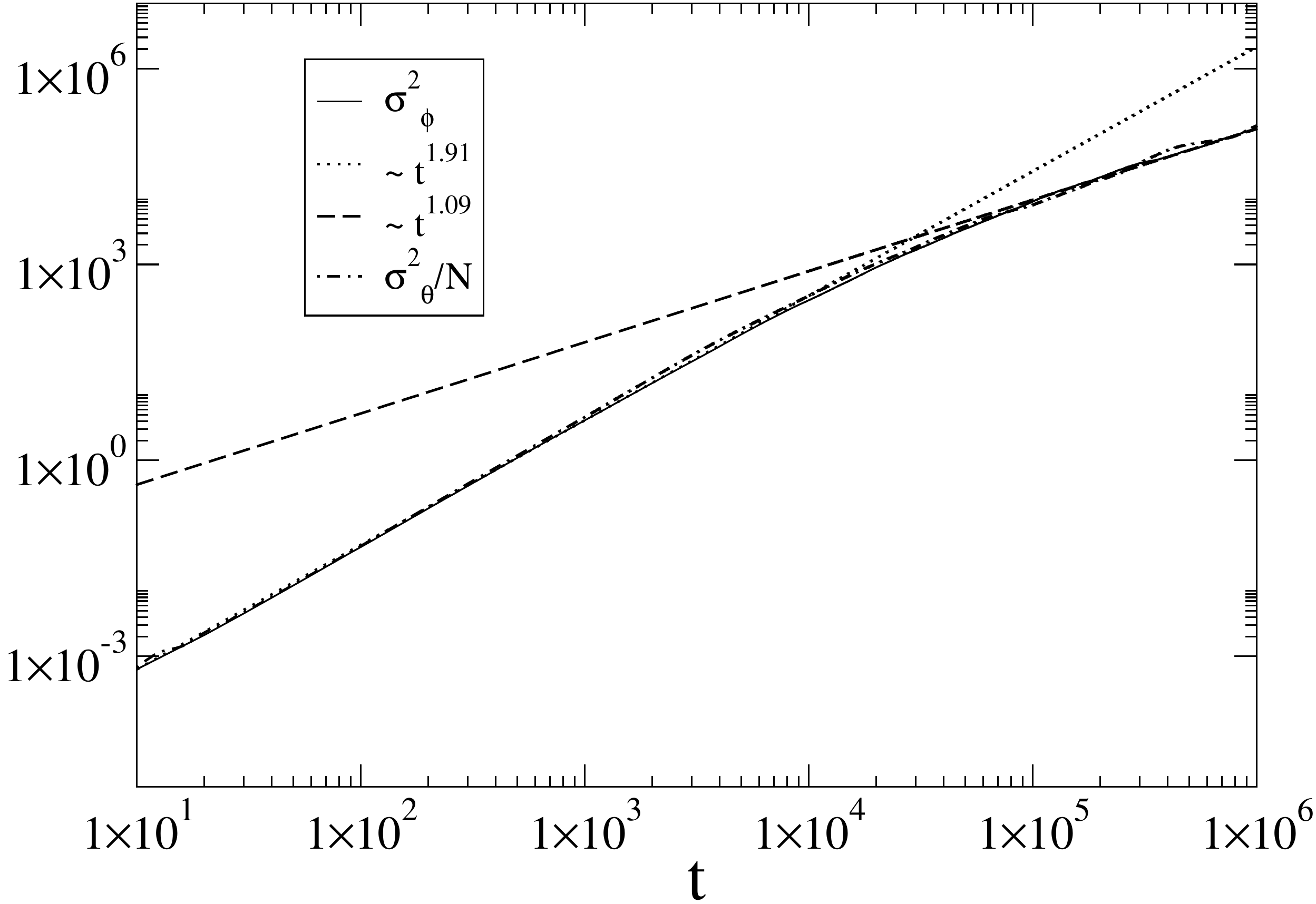}}}
\end{center}
	\caption{Variance $\sigma^2_\phi(t)$ for the position of the center of mass at equilibrium for an equilibrium state of
the HMF model with $e=0.4$, $N=5000$ and $500$ realizations.
The variance $\sigma^2_\theta(t)$ for the position variables of each individual particle is also shown rescaled
	by the number of particles $N$ which collapses to the values of $\sigma^2_\phi(t)$, in agreement with Eq.~(\ref{difcmdiffpart2}).}
\label{alphaMD}
\end{figure}

\begin{figure}[ptb]
\begin{center}
\scalebox{0.3}{{\includegraphics{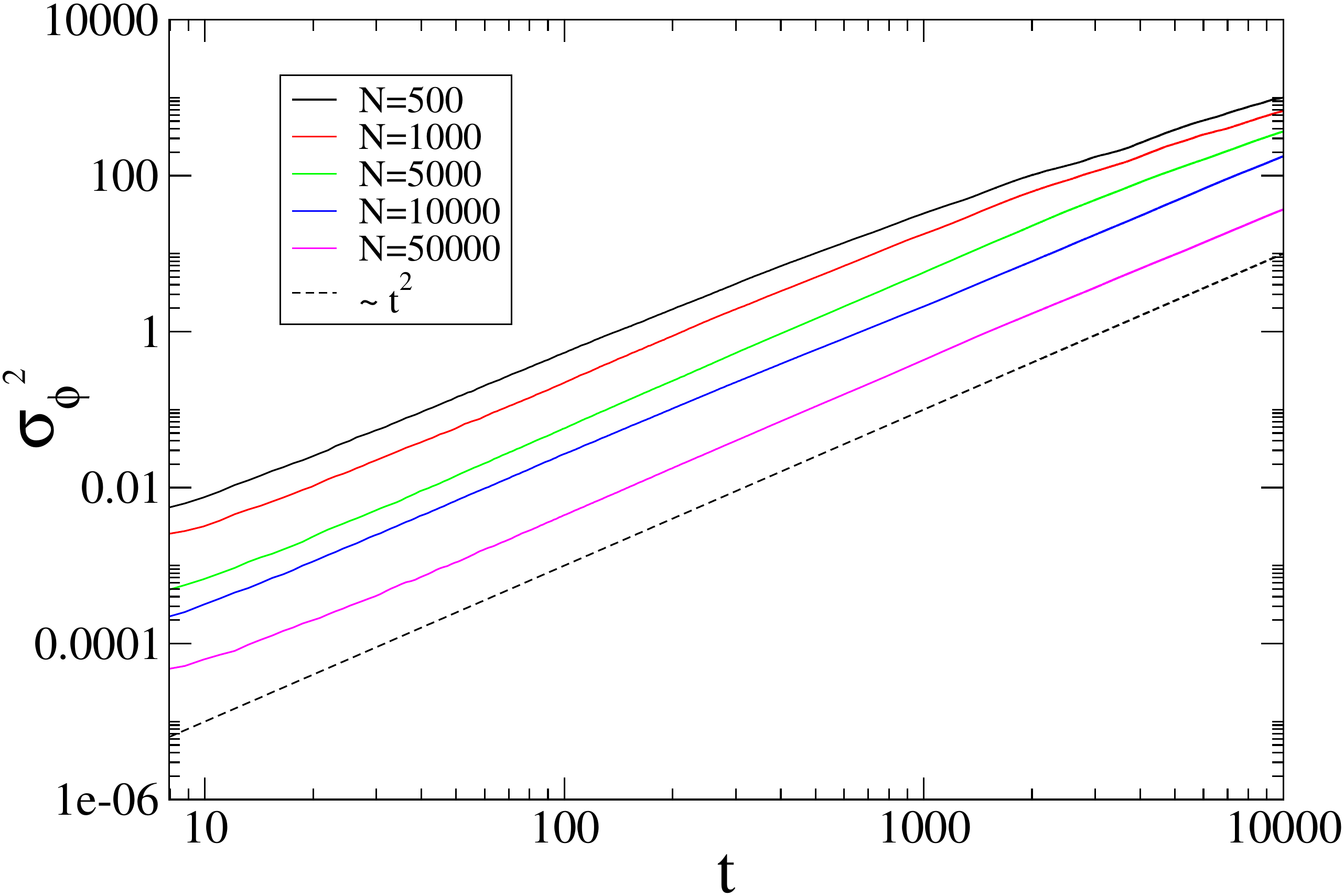}}}
\scalebox{0.3}{{\includegraphics{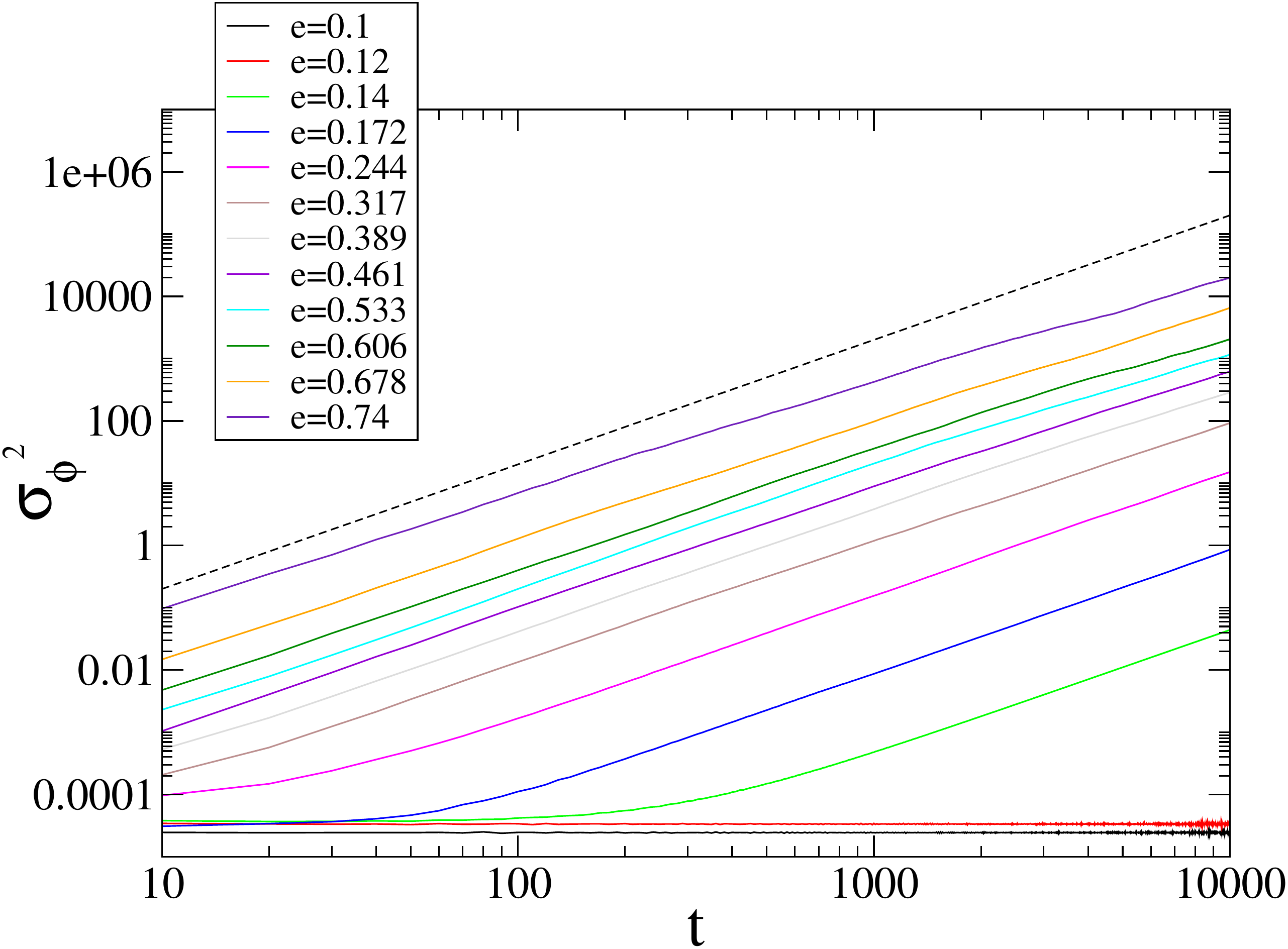}}}
\end{center}
	\caption{(Color online) Left panel: Variance $\sigma^2_\phi(t)$ for the position of the center of mass at equilibrium for an equilibrium state of
        the HMF model with $e=0.4$, $100$ realizations and a few values of $N$. The dashed line is proportional to $t^2$ and is given for
        comparison purposes. Right panel: Variance $\sigma^2_\phi(t)$ for the equilibrium state for $N=5000$ and different values of
	energy per particle. For very low energies there is almost no diffusion, as expected.}
\label{alphaMDns}
\end{figure}

In order to characterize the diffusive movement of the center of mass of the HMF model
we compute the square root displacement $\sigma_\phi(t)=\sqrt{\langle\phi(t)^2\rangle}$, with $\langle\phi(t)\rangle=0$
(recall that $\phi$ is defined in the extended space i.~e.\ $\phi\in(-\infty,\infty)$). Figure~\ref{alphaMD} shows the results
for $e=0.4$ and $N=5000$. A power law fit for the initial and final parts of the plot,
shows that the motion is initially superdiffusive close to ballistic and tends to normal diffusion asymptotically.
The variance of individual particle position $\sigma^2_\theta(t)$ is also shown in the figure rescaled by a factor $N$,
showing a very good agreement with Eq.~(\ref{difcmdiffpart2}).
Figure~\ref{alphaMDns} shows the variance $\sigma_\phi^2$ as a function of time for different values of $N$ and fixed energy
(left panel), and different values of the energy per particle $e$ for $N=5000$. The diffusion is close to ballistic
for the time window considered, and tends to disappear for lower energies as the probability of a particle to reach the boundary
of the physical space (with respect to the peak of the distribution) goes to zero as $e\rightarrow0$.

The anomalous diffusion of particles in the HMF model and the
periodic boundary conditions translate into an anomalous diffusion of the center of mass of the whole system.
As commented above, anomalous diffusion in the HMF model was studied by some authors~\cite{difus1,latora,bouchet,yamaguchi2,chavanis3,moyano},
and superdiffusion was shown to be a common feature, even at equilibrium.

\subsection{Dynamics of the center of mass}

For the HMF model a complete theoretical characterization of the initial ballistic diffusive motion of the center of mass is possible.
We consider here the case of the equilibrium state but the approach can be easily generalized for more
general (quasi-) stationary states. We first characterize the jumps of the position of the center of mass by showing that is
given by the difference of two Poisson processes. Then we discuss how to compute the coefficient of the initial ballistic diffusion
and why it tends to normal diffusion due to finite $N$ effects, i.~e.\ collisions or granularity effects.

\subsection{Statistics of the center of mass jumps}

\begin{figure}[ptb]
\begin{center}
\scalebox{0.3}{{\includegraphics{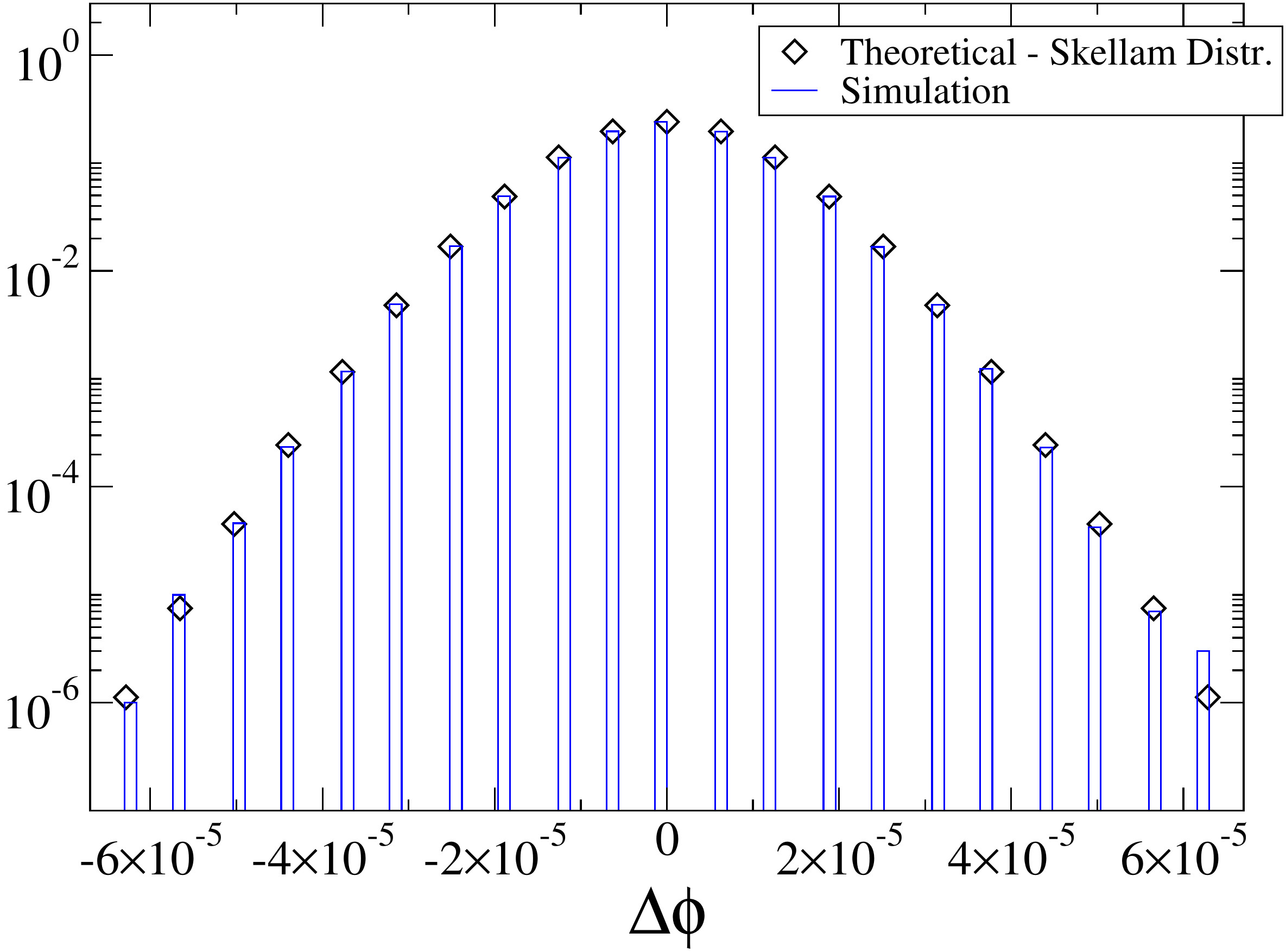}}}
\end{center}
	\caption{(Color online) Normalized histograms (vertical bars) from a numeric simulation for the frequency of increments $\Delta\phi$
of the center of mass position recorded after each time step $\Delta t=0.01$, total simulation time $t_f=10^5$,
energy $e=0.4$ and $N=1000\,000$ compared to the distribution in Eq.~(\ref{skellamdistphi}) (diamonds).}
\label{Dalpha}
\end{figure}

Let us consider the equilibrium one-particle distribution function given in Eq.~(\ref{hmfeqstate}), initially centered at
$\theta=0$ ($M_y=0$ and $M_x=M$). The probability that a given particle crosses at $\theta=\pi$ with $p>0$
during a small time interval $\Delta t$ is given by:
\begin{equation}
{\cal P}_+=\int_0^\infty\dd p\int_{\pi-p\Delta t}^\pi\dd\theta\:f_{\rm eq}(\theta,p)=
\frac{e^{-\beta M}\Delta t}{(2\pi)^{3/2}\sqrt{\beta}I_0(\beta M)},
\label{resPplus}
\end{equation}
and the probability that a given particle traverses at $\theta=-\pi$ with $p<0$ is
\begin{equation}
{\cal P}_-=\int_0^\infty\dd p\int_{-\pi}^{-\pi+p\Delta t}\dd\theta\:f_{\rm eq}(\theta,p)=
\frac{e^{-\beta M}\Delta t}{(2\pi)^{3/2}\sqrt{\beta}I_0(\beta M)},
\label{resPminus}
\end{equation}
which is, obviously, the same as ${\cal P}_+$. Thus the probability that one particle, no matter which, crosses at each one of the boundaries
at $\theta=\pm\pi$ is ${\cal P}=N{\cal P}_+$. Now supposing that for sufficiently small $\Delta t$ the crossings of particles
are independent from each other, the probability that $\Delta N$ particles cross at one of the boundaries is given by the
Poisson distribution:
\begin{equation}
P(\Delta N)=e^{-\cal P}\frac{{\cal P}^{\Delta N}}{\Delta N!}.
\label{poissondist}
\end{equation}
The probability for the value of the difference $c=a-b$ of two Poisson distributed random variables $a$ and $b$, with respective averages
$\overline{a}$ and $\overline{b}$, is given by the Skellam distribution~\cite{skellam}:
\begin{equation}
{\cal S}(c)=e^{-(\overline{a}+\overline{b})}\left(\frac{\overline{a}}{\overline{b}}\right)^cI_{c}\left(2\sqrt{\overline{a}\overline{b}}\right),
\label{defskellamgen}
\end{equation}
with $I_c$ a modified Bessel function with index $c$.
Now considering that $\Delta N_+$ and $\Delta N_-$ particles cross at $\theta=\pi$ and $\theta=-\pi$, respectively,
in the time interval $\Delta t$, and noting that $\overline{a}=\overline{b}={\cal P}$,
the probability that the difference, i.~e.\ the net flux, is $\Delta N=\Delta N_+-\Delta N_-$ is given by:
\begin{equation}
{\cal S}(\Delta N)=e^{-2{\cal P}}I_{|\Delta N|}\left(2{\cal P}\right).
\label{skellamdist}
\end{equation}
For a given net flux of particles at the border $\Delta N$, the center of mass moves by $\Delta\phi=-2\pi\Delta N/N$. Hence
the probability that the center of mass moves by $\Delta\phi$ in the same time interval $\Delta t$ is:
\begin{equation}
{\cal S}(\Delta\phi)=e^{-2{\cal P}}I_{|N\Delta\phi/2\pi|}(2{\cal P}).
\label{skellamdistphi}
\end{equation}
Since the possible values of $\Delta\phi$ are discrete there is no extra multiplication factor
resulting from going from Eq.~(\ref{skellamdist}) to Eq.~(\ref{skellamdistphi}).
Figure~\ref{Dalpha} shows the frequencies (histograms) of $\Delta\phi$ obtained from a very long run and the theoretical distribution
in Eq.~(\ref{skellamdistphi}) with a very good agreement.
For $\Delta N$ large, the Skellam distribution tends to a Gaussian distribution of the form~\cite{skellam}:
\begin{equation}
{\cal S}(\Delta\phi)\rightarrow\frac{N}{2\pi^{3/2}\sqrt{\cal P}}\:\exp\left(-\frac{N^2\Delta\phi^2}{4\pi^2{\cal P}}\right).
\label{skellgauss}
\end{equation}

We will see in the next sections that the statistics of the jumps is {\em not} sufficient to fully characterize the diffusion process.
Time-correlation in the jumps are very important, as we will detail below.

\subsection{The variance of the position of the center of mass $\phi$}
The variance of the position of the center of mass of the system
is written as:
\begin{eqnarray}
\sigma^2_\phi(t) & = & \langle[\phi(t)]^2\rangle=
\left\langle\frac{1}{N}\sum_{i=1}^N\theta_i(t)\times\frac{1}{N}\sum_{i=j}^N\theta_j(t)\right\rangle
\nonumber\\
 & = & \frac{1}{N^2}\sum_{i,j=1}^N\left\langle\int_0^t{\rm d}t^\prime p_i(t^\prime)
\int_0^t{\rm d}t^{\prime\prime} p_j(t^{\prime\prime})\right\rangle
\nonumber\\
 & = & \frac{t}{N}\int_0^t{\rm d}\tau\:{\cal C}_p(\tau),
\label{variancephi}
\end{eqnarray}
where we used the property
${\cal C}_p\equiv\left\langle p(0)p(\tau)\right\rangle=\left\langle p(t)p(t+\tau)\right\rangle$,
valid for a stationary state. In function of the convergence properties of ${\cal C}_p$ in Eq.~\eqref{variancephi}, the center of mass $\phi$ will experiment  ballistic or normal diffusion.

\subsection{Ballistic diffusion}
\label{subsect-ballistic}
Long-term memory of the initial condition is a characteristic property of systems with long-range interactions,
and one consequence is anomalous diffusion~\cite{fernando}. 
The ballistic initial diffusion of the center of mass can be explained by the
fact that, for a mean-field system, the momentum auto-correlation function tends to zero after a collisional characteristic time $\tau_{coll}$,
which is the time interval collisional effects destroy the memory of the initial state. It is well known that in spatially inhomogeneous
configurations of the HMF system, $\tau_{coll}$ scales linearly with $N$~\cite{balescu,scaling,scaling2}. In particular,
in the limit $N\rightarrow\infty$, the momentum auto-correlation never vanishes.

\begin{figure}[ht]
\scalebox{0.3}{{\includegraphics{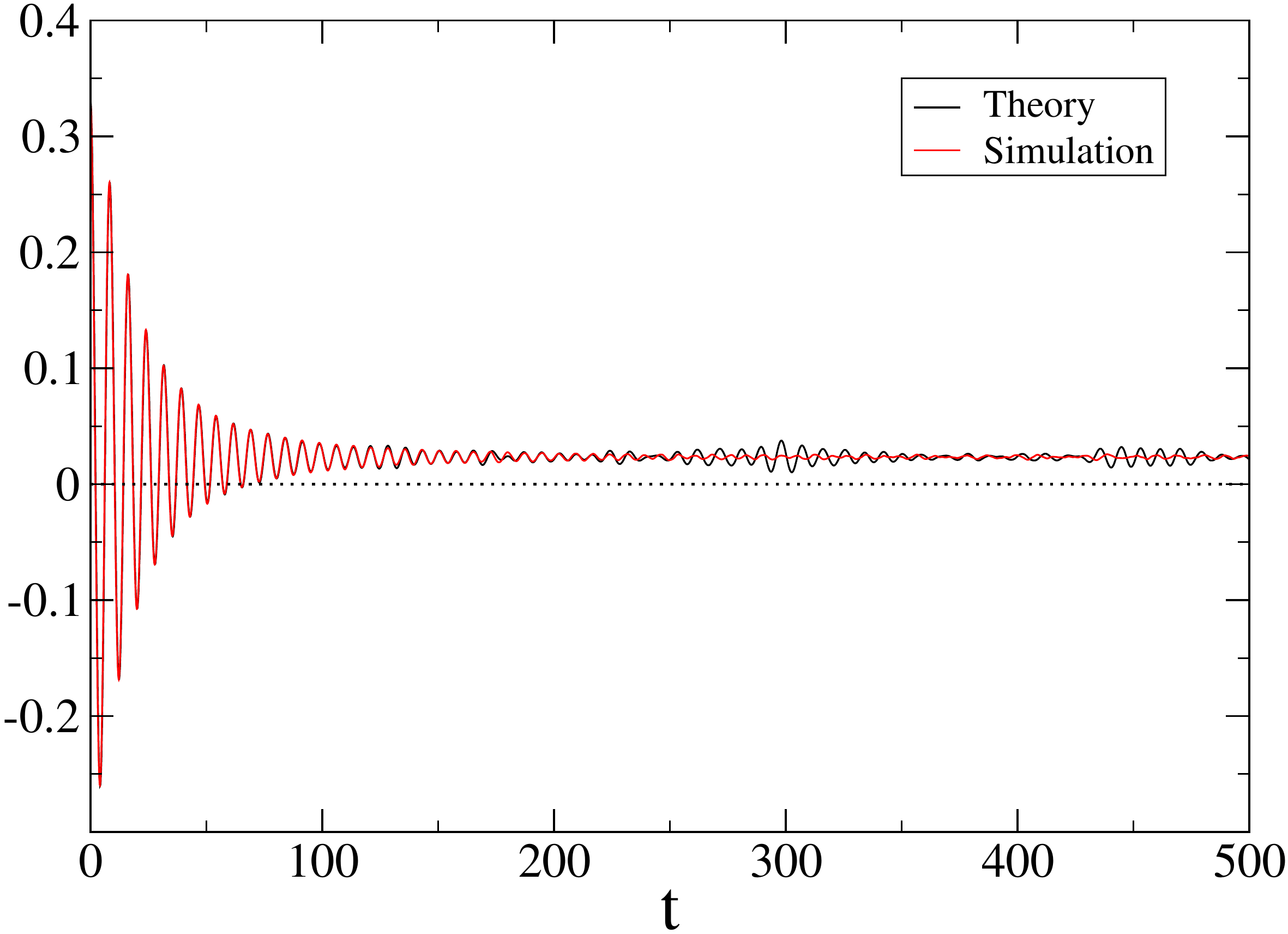}}}
\caption{(Color online) Momentum auto-correlation function ${\cal C}_p(t)$ obtained from Eq.~(\ref{Cpdeterm}) and from a numeric simulation
for $e=0.4$ and $N=1000\,000$. The dotted line was introduced for reference. We see that
${\cal C}_p(t)$ tends asymptotically to a non-vanishing value.}
\label{autocorrelp}
\end{figure}

In a stationary state in the thermodynamic limit the motion of a particle obeys the equations of a pendulum:
\begin{equation}
\dot\theta=p,\hspace{3mm}\dot p=-M\sin(\theta),
\label{eqmotmflim}
\end{equation}
with known closed form solution in terms of an elliptic function for initial conditions $\theta(0)=\theta_0$ and $p(0)=p_0$, and therefore
the auto-correlation function ${\cal C}_p$ for this stationary state can be determined exactly (up to two integrations) as:
\begin{equation}
{\cal C}_p(\tau)=\int_{-\infty}^\infty\dd p_0\int_{-\pi}^\pi\dd\theta_0\:f_{\rm st}(\theta_0,p_0)\:p_0\:p(\tau),
\label{Cpdeterm}
\end{equation}
which is valid for time $t\ll \tau_{coll}$ and
where $f_{\rm st}$ denotes the one-particle distribution function for the stationary state. For the equilibrium state
$f_{st}$ is given by Eq.~(\ref{hmfeqstate}) and and $p(t)$ is the solution of the equation
\begin{equation}
	Q(p(t))-Q(p_0)=t,
	\label{solptau}
\end{equation}
with
\begin{eqnarray}
	Q(p) \equiv \pm \sqrt{2}\:\frac{\sin(p/2)}{\sqrt{e-M}}{\cal F}\left(\cos(p/2),\sqrt{\frac{2M}{M-e}}\right),
	\label{solpendulum}
\end{eqnarray}
where ${\cal F}$ is the incomplete elliptic integral of the first kind.
The plus and minus sign in the right-hand side of Eq.~(\ref{solpendulum}) represent the two different branches of the solution.
An easy way to overcome the analytical computation of the resulting cumbersome integral in Eq.~(\ref{Cpdeterm}) is to compute it numerically with
any desired accuracy and a small numeric effort.
Figure~\ref{autocorrelp} shows the auto-correlation function at equilibrium for $e=0.4$ obtained from Eq.~(\ref{Cpdeterm}),
and the same function obtained from a fully numeric molecular dynamics simulation, with a very good agreement.
We see that for $t\ll \tau_{coll}$, or equivalently in the limit $N\rightarrow\infty$ for any time, the correlation function
takes a non-vanishing value $\tilde{\cal C}_p$. Using Eq.~\eqref{variancephi} the variance of position of the center of mass is then
\begin{equation}
\sigma^2_\phi(t)=\frac{t}{N}\int_0^t d\tau\: \tilde{\cal C}_p=\frac{\tilde{\cal C}_p}{N}\:t^2\equiv\sigma_N^2\:t^2.
	\label{varCp}
\end{equation}
This explains why the diffusion is initially ballistic, or close to ballistic for $t\ll \tau_{coll}$.
In Fig.~\ref{autocorrelp}, we can see that it is indeed the case. After a transient between $t=0$ and $t\approx 200$,
the momentum auto-correlation function takes a constant value.
By replacing $f_{\rm eq}$ in the above expression for any stationary state, all results above remain valid.

The value of the constant $\sigma_N^2$ can be obtained explicitly using the fact that the one-particle phase space is divided by a
separatrix for points corresponding to a libration (outside the separatrix), and bounded motion (inside the separatrix).
The separatrix is defined such that the one-particle energy equals the maximum of the mean-field potential. The particles
which contribute to the ballistic diffusion are those which are librating, i.e. outside the separatrix. This is because
the positions of the particles which are outside the separatrix can increase indefinitely whereas this is not the case
for those which lie inside the separatrix. We can therefore write, after a transient time, the position of the center of mass as
\begin{equation}
\phi \simeq \frac{1}{N}\sum_{i=1}^{N^+}\theta_i^+(t),
\end{equation}
where $\theta^+$ are the $N^+$ particles which lie outside the separatrix, and thus
\begin{equation}
\langle\phi^2\rangle \simeq  \frac{1}{N}  \langle (\theta^+)^2 \rangle \simeq \frac{1}{N}  \langle (v^+)^2 \rangle t^2,
\end{equation}
where $\langle (v^+)^2 \rangle$ is the variance of the velocity of the particles outside the separatrix. We have therefore
\begin{equation}
\label{sigma-velplus}
\sigma_N^2 \simeq   \langle(v^+)^2 \rangle.
\end{equation}

Note that, as the system is at equilibrium, the quantity $\langle (v^+)^2 \rangle$  does not depend on time. 
We need first to compute the velocity distribution of the particles with an energy
larger than the separatrix, which we will call $P^+(v)$. For a system
with an average magnetization $M$, particles are outside the
separatrix if their energy $e$ is larger than the average
magnetization, i.e.,
\begin{equation}
e=\frac{v^2}{2}-M\cos\theta \ge M,
\end{equation}
where we have used without loss of generality that $M_y=0$ and then $M=M_x$.  
The first step in the calculation is to compute the probability density of $\cos\theta$. Using the equilibrium distribution function
in Eq.~(\ref{hmfeqstate}) we get
\begin{equation}
P\left(X=\cos\theta\right)=\int_0^{2\pi}d\theta \frac{\exp(\beta M \cos\theta)}{2\pi I_0(\beta M)}\delta(X-\cos\theta)=\frac{1}{\pi I_0(\beta M)}\frac{\exp(\beta M X)}{\sqrt{1-X^2}}.
\end{equation}
We are interested in the probability
\begin{equation}
P\left(-1\le\cos\theta\le \frac{v^2}{2M}-1\right)\equiv F(v,\beta)=\frac{1}{\pi I_0(\beta M)}\int_{-1}^{\frac{v^2}{2M}-1} dX \frac{\exp(\beta M X)}{\sqrt{1-X^2}}.
\end{equation}
The integral in this equation cannot be performed analytically.

\begin{figure}[ptb]
\begin{center}
\scalebox{1.1}{{\includegraphics{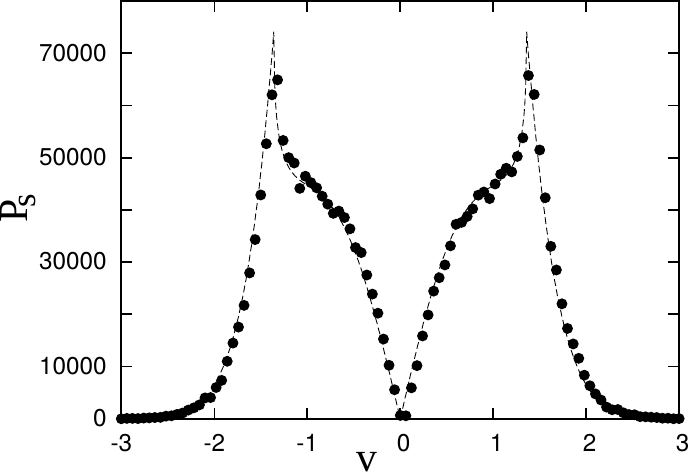}}}
\end{center}
\caption{Comparison of the distribution in Eq.(\ref{distri-theo}) (dashed line) and a numerical
realization (circles) with $N=10^8$ particles and $\beta=2.26$.}
\label{distrifig}
\end{figure}

There are two possible cases according to the velocity of the particles:
\begin{enumerate}
\item if $|v|>2\sqrt{M}$, then the particle automatically lies outside the separatrix.
\item if $|v|< 2\sqrt{M}$, then the particle is outside the separatrix only if $\cos\theta < v^2/2M-1$.
\end{enumerate}
The velocity distribution of the particles outside the separatrix is thence:
\begin{equation}
\label{distri-theo}
P^+(v,\beta) = \begin{cases} \sqrt{\frac{\beta}{2\pi}}\exp(-\beta v^2/2), &\mbox{if }|v|>2\sqrt{M}.
\\
\sqrt{\frac{\beta}{2\pi}}\exp(-\beta v^2/2)F(v,\beta), &\mbox{if }|v|< 2\sqrt{M}.
\end{cases} 
\end{equation}
The distribution in Eq.~(\ref{distri-theo}) is shown in Fig.~\ref{distrifig} with a comparison to
a numerical realization with $N=10^6$ particles.

We compute now the variance of the velocity of the particles outside the separatrix:
\begin{equation}
\label{sigma-ballistic}
\langle (v^+)^2 \rangle= \int_{-\infty}^{\infty} dv\: v^2 P^+(v,\beta).
\end{equation}
Using Eq.~\eqref{distri-theo}, we get to the contribution of the integral for $|v|> 2\sqrt{M}$:
\begin{equation}
\label{approx1}
2\int_{2\sqrt{M}}^{\infty} dv\: v^2 P^+(v,\beta) = 2\sqrt{\frac{2M}{\pi \beta}}+\mathrm{Erfc}\left( \sqrt{2\beta M}\right).
\end{equation}
 For sufficiently large $\beta$ (i.e. not too close to the phase transition $\beta=2$), and using that, for these values of $\beta$, 
\begin{equation}
M\simeq 1-\frac{1}{2\beta}+\mathcal{O}\left(1/\beta^2\right),
\end{equation}
this expression can be approximated with
\begin{equation}
2\int_{2\sqrt{M}}^{\infty} dv\: v^2 P^+(v,\beta)= 2\sqrt{\frac{2}{\pi\beta}}e^{1-2\beta +\mathcal{O}\left(1/\beta\right)}.
\end{equation}
To get an analytic approximation of the contribution of integral \eqref{sigma-ballistic} for $|v|> 2\sqrt{M}$ it is convenient to invert the order of integration between $x$ and $v$. We get 
\begin{equation}
\label{diff_part}
2\int_{0}^{2\sqrt{M}} dv\: v^2 P^+(v,\beta) =\frac{1}{\pi I_0(\beta M)}\int_{-1}^1 dx\: \frac{e^{\beta M x}}{\sqrt{1-x^2}}g(x,\beta)
\end{equation}
where 
\begin{equation}
\label{g-function}
g(x,\beta)=\frac{\text{Erf}\left( \sqrt{2\beta M}\right)+\text{Erfc}\left(\sqrt{\beta M (x+1)}\right)+\frac{2 \left(e^{-\beta M (x+1)} \sqrt{\beta M (x+1)}-\sqrt{2} e^{-2 \beta M} \sqrt{\beta M}\right)}{\sqrt{\pi }}-1}{2 \beta}.
\end{equation}
Since integral \eqref{diff_part} is dominated by the region $x\sim 1$, in order to get an analytical approximation, it is possible to expand the function $\text{Erfc}\left(\sqrt{b M (x+1)}\right)$ in power series around $x=1$. It is then  possible to find an analytical expression for Eq.~\eqref{diff_part}, which is, for sufficiently large $\beta$:
\begin{equation}
\label{approx2}
2\int_{0}^{2\sqrt{M}} dv\: v^2 P^+(v,\beta)F(v,\beta) = \left(\frac{8}{\pi}-\frac{33}{8\sqrt{2\pi \beta}}+\mathcal{O}\left(\frac{1}{\b}\right)\right)e^{1-2\beta+\mathcal{O}(1/\b)}.
\end{equation}
Combining Eqs.~\eqref{approx1} and \eqref{approx2} we obtain that, at leading order
\begin{equation}
	\sigma_N^2=\frac{\tilde{C}_p}{N}\simeq\frac{8}{\pi}e^{1-2\beta}.
	\label{valCp}
\end{equation}

\begin{figure}
\scalebox{0.3}{{\includegraphics{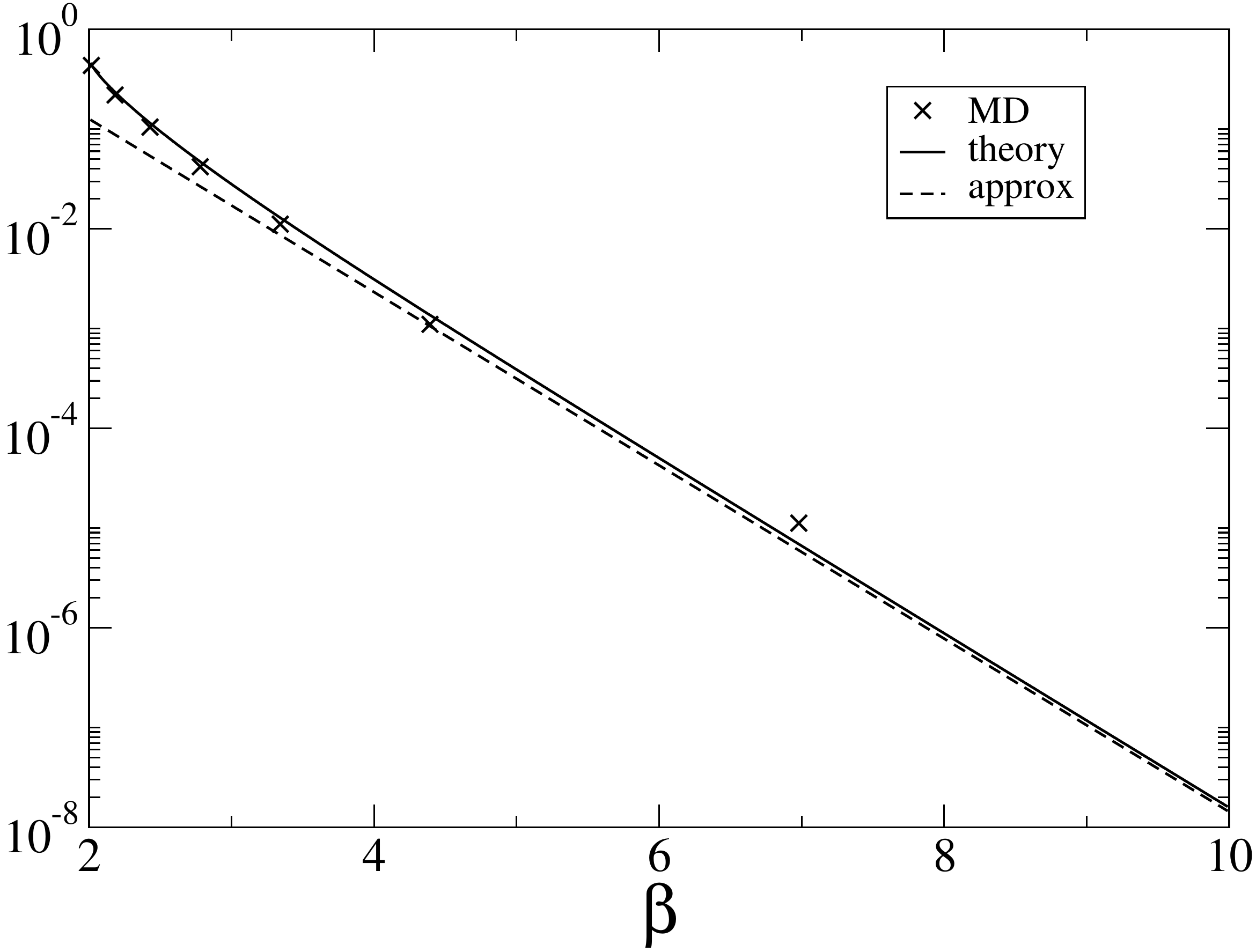}}}
\scalebox{0.3}{{\includegraphics{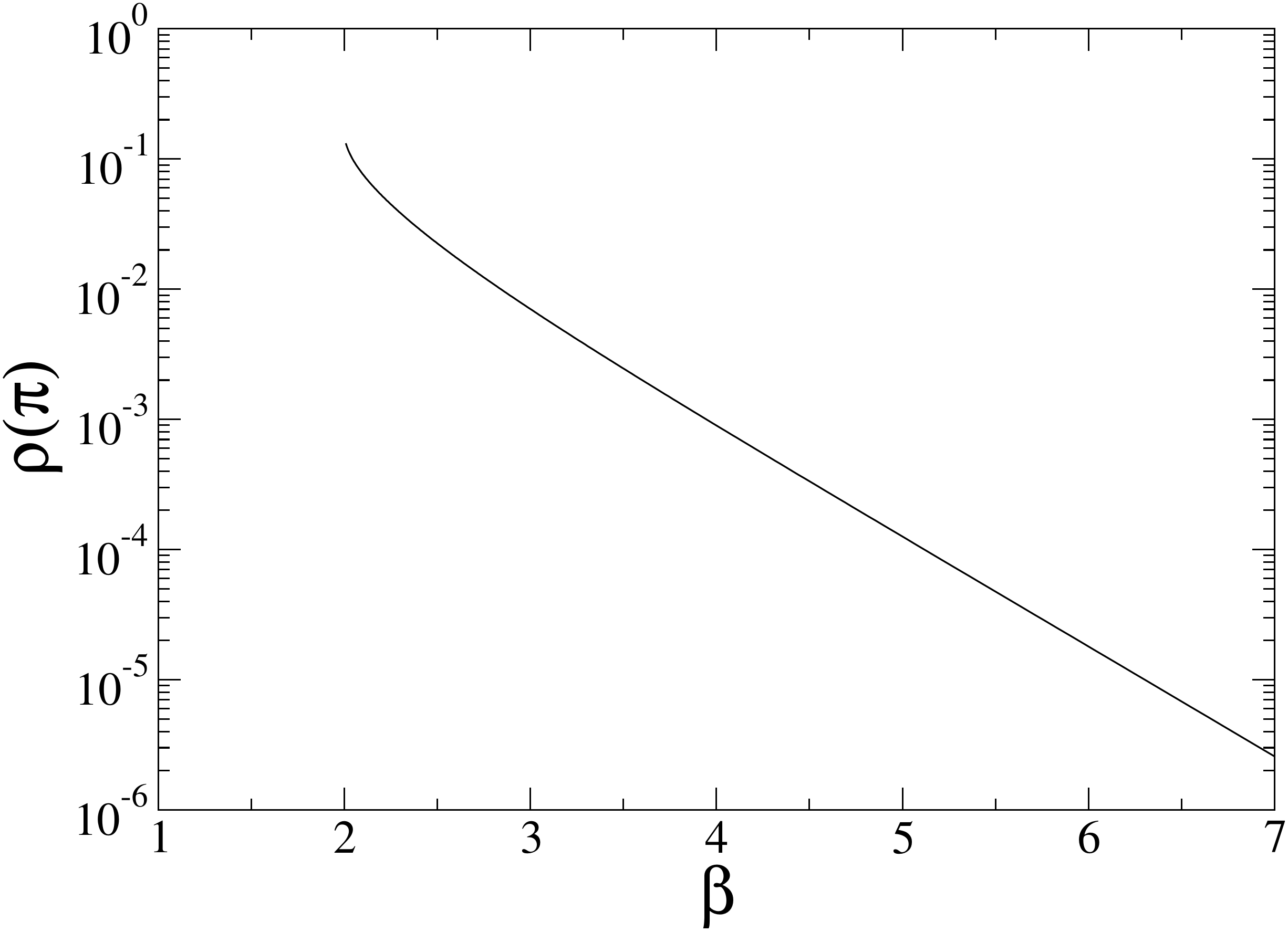}}}
        \caption{Left panel: Ballistic diffusion coefficient $\tilde{\cal C}_p$ from Eq.~(\ref{Cpdeterm}), molecular dynamics (MD) simulations,  theoretical prediction Eqs.~\eqref{sigma-velplus} and (\ref{sigma-ballistic}), and analytical approximation \eqref{valCp}. Right panel: Spatial distribution function
	at $\theta=\pi$ from Eq.~(\ref{distri-space}). We see that $\sigma_N^2$
is roughly proportional to $\rho(\pi)$ when the center of mass is located at the origin, as expected,
	and the flow of particles is proportional to $\rho(\pi)$.}
\label{diffcompar}
\end{figure}

A comparison of $\tilde{\cal C}_p$ obtained from Eq.~(\ref{sigma-ballistic}) with numeric simulations for different values of $\beta$
is shown in the left-panel of Fig.~\ref{diffcompar} with a good very agreement.
The spatial distribution function obtained
using Eq.~(\ref{hmfeqstate}) is:
\begin{equation}
\label{distri-space}
\rho(\theta,t)=\frac{1}{2\pi I_0(\beta M)} e^{\beta M \cos(\theta+\phi(t))},
\end{equation}
and is shown on the right-panel of the same figure. From Eq.~(\ref{resPplus}) we have that the number of particles that
cross at the boundary at $\theta=\pi$ during the time interval $\Delta t$ is thus given by
\begin{equation}
	{\cal P}_+=\frac{\Delta t}{\sqrt{2\pi\beta}}\rho(\pi).
	\label{relPrho}
\end{equation}
We see that $\sigma_N^2$ is roughly proportional to $\rho(\pi)$, the value of the spatial density at $\theta=\pi$ for $\phi=0$.
This illustrates the fact that the diffusive ballistic motion is indeed due to an excess of particles crossing at the boundaries
into different directions at the boundary of the periodic variable $\theta$.

\subsection{Normal diffusive regime}

\begin{figure}[ht]
\scalebox{0.3}{{\includegraphics{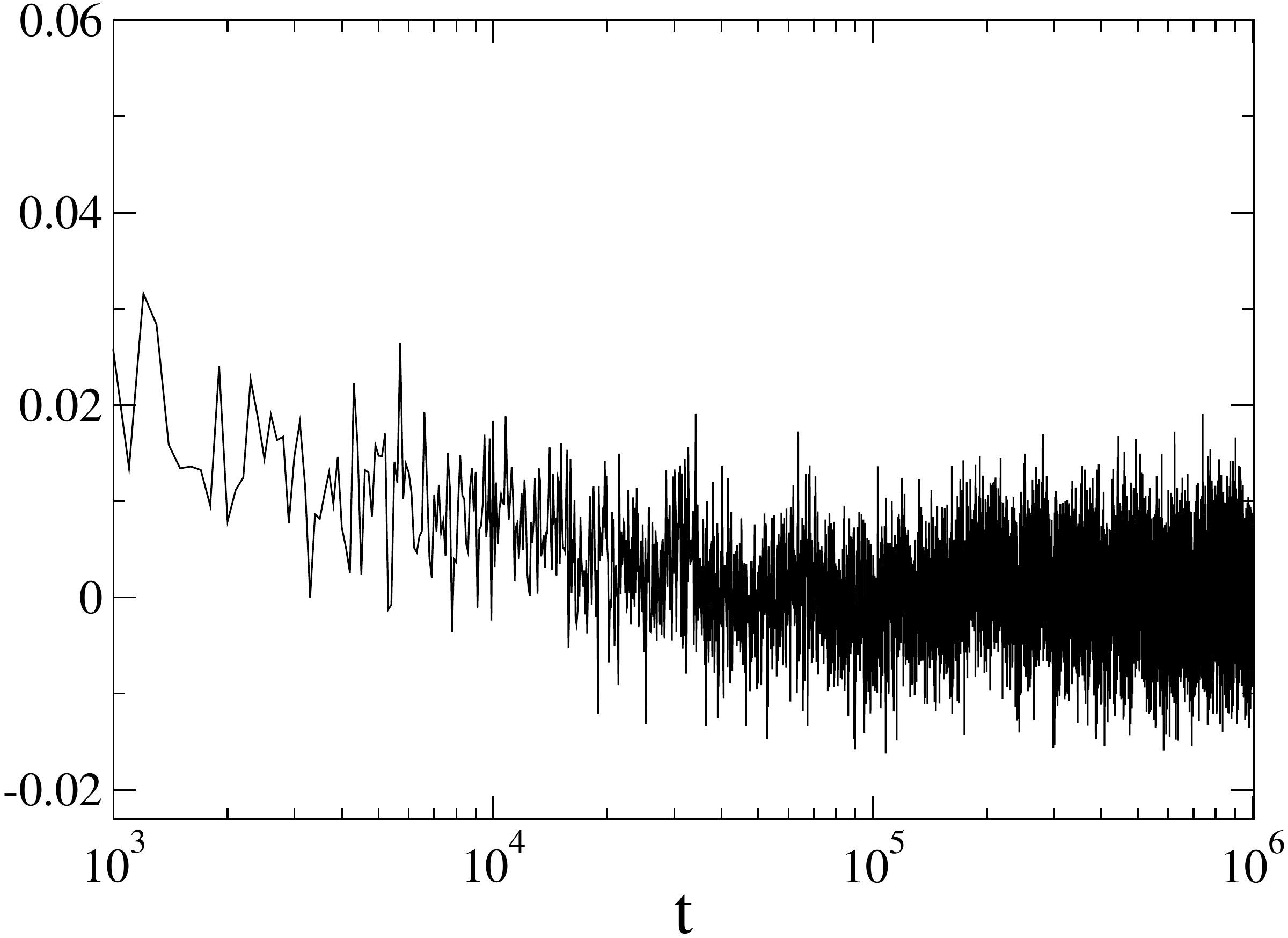}}}
\caption{Momentum auto-correlation function ${\cal C}_p(t)$ at equilibrium for very long times,
$e=0.4$, $N=10\,000$. Note that the time required for ${\cal C}_p$
to reach zero corresponds to the crossover time from non-normal to normal diffusion in Fig.~\ref{alphaMD}}
\label{autocorrelplong}
\end{figure}

\begin{figure}[ht]
\scalebox{1.1}{{\includegraphics{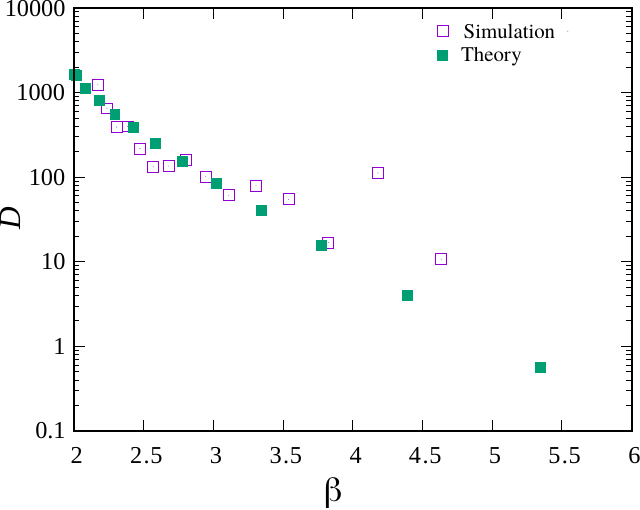}}}
	\caption{Normal diffusion coefficient $D$ at equilibrium of the center of mass as a function of $\beta$. The simulation has been performed for $N=1000$, $50$ realizations
	and total simulation time $t_f=10^6$.}
\label{normaldiffcoeff}
\end{figure}

For finite $N$, collisional effects destroy the memory of the initial state on a time scale proportional to
the order of the strength of the interaction, which for non-homogeneous states is $1/N$~\cite{balescu,scaling,scaling2},
causing the auto-correlation function to slowly approach zero, as exemplified in Fig.~\ref{autocorrelplong}.
Consequently the diffusion tends to normal in this same time scale, after which the variance of the center of mass
position satisfies $\sigma_\phi(t)^2=D\cdot t$, with $D$ the (normal) diffusion coefficient.
The precise theoretical determination of the crossover time between anomalous and normal diffusion and the value of $D$ is a very difficult task
in kinetic theory, and well beyond the scope of the present work. We can however determine the diffusion coefficient using an
approximation for the exact expression for the variance of position of the center of mass:
\begin{equation}
\label{var_normal}
	\sigma^2_\phi(t)=\frac{t}{N}\int_0^\infty d\tau\: {\cal C}_p(\tau).
\end{equation}
We know that the correlation coefficient has the form
\begin{equation}
{\cal C}_p(\tau) = \tilde{\cal C}_p f(\tau,\beta),
\end{equation}
where $f(\tau,\beta)$ is an unknown function of time and $\beta$ related to the collisional relaxation process with $f(0,\beta)=1$,
$f(\tau\to\infty,\beta)=0$ and $\tilde{\cal C}_p$ defined in Eq.~(\ref{varCp}). This  describes the behavior of the correlation function observed
in Fig.~\ref{autocorrelplong} for a particular value of $\beta$. If we assume that the function $f$ does not depend strongly on $\beta$ we can write
\begin{equation}
{\cal C}_p(\tau) \simeq \tilde{\cal C}_p f(\tau),
\end{equation}
and thence for the variance of position of the center of mass:
\begin{equation}
\label{var_normal-app}
\sigma^2_\phi(t)\simeq\frac{t}{N}\int_0^\infty d\tau\tilde{\cal C}_p f(\tau) = \frac{\tilde{\cal C}_p}{N}\,t\int_0^\infty d\tau f(\tau).
\end{equation}
We compute numerically the last integral in the right-hand side of Eq.~(\ref{var_normal-app}) for $e=0.4$, obtaining
\begin{equation}
\int_0^\infty d\tau f(\tau)\approx 730.
\end{equation}
Using this result and the analytical expression for $\tilde{\cal C}_p$ in Eq.~(\ref{valCp}) we show in Fig.~\ref{normaldiffcoeff} the
normal diffusion coefficient $D$ a function of $\beta$ with a  good agreement between theory and simulation. 
Note that to obtain the numerical estimate requires a considerable numeric effort with very long integration times, and with the caveat
that the higher the value of $N$ the higher the crossover time. As expected, $D$ tends to zero for decreasing energy (increasing $\beta$).

\section{Classical Goldstone modes and chaos}
\label{sec6}

In nematic liquid crystals the coupling of a roll pattern of electroconvection with a Goldstone mode, due
to the symmetry breaking of the alignment of the nematic molecules, results in what is known as
soft-mode turbulence~\cite{hidaka}. We show now that, similarly, the coupling of the thermal excitations of a Goldstone mode,
related to a periodic coordinate in long-range systems, to the mean-field motion
of the particles, may lead to what is called strong chaotic behavior.

\begin{figure}
\begin{center}
\scalebox{0.3}{{\includegraphics{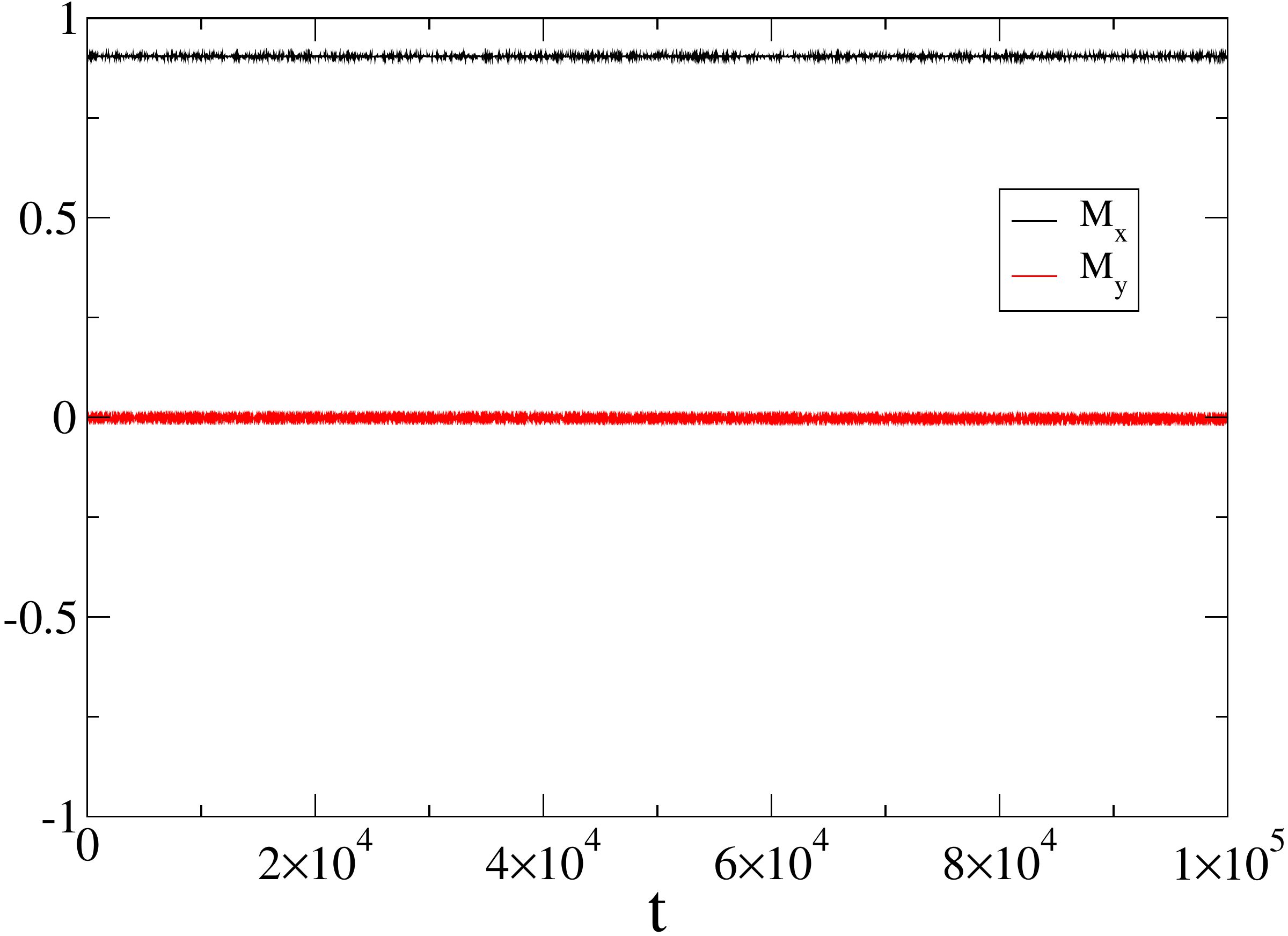}}}
\scalebox{0.3}{{\includegraphics{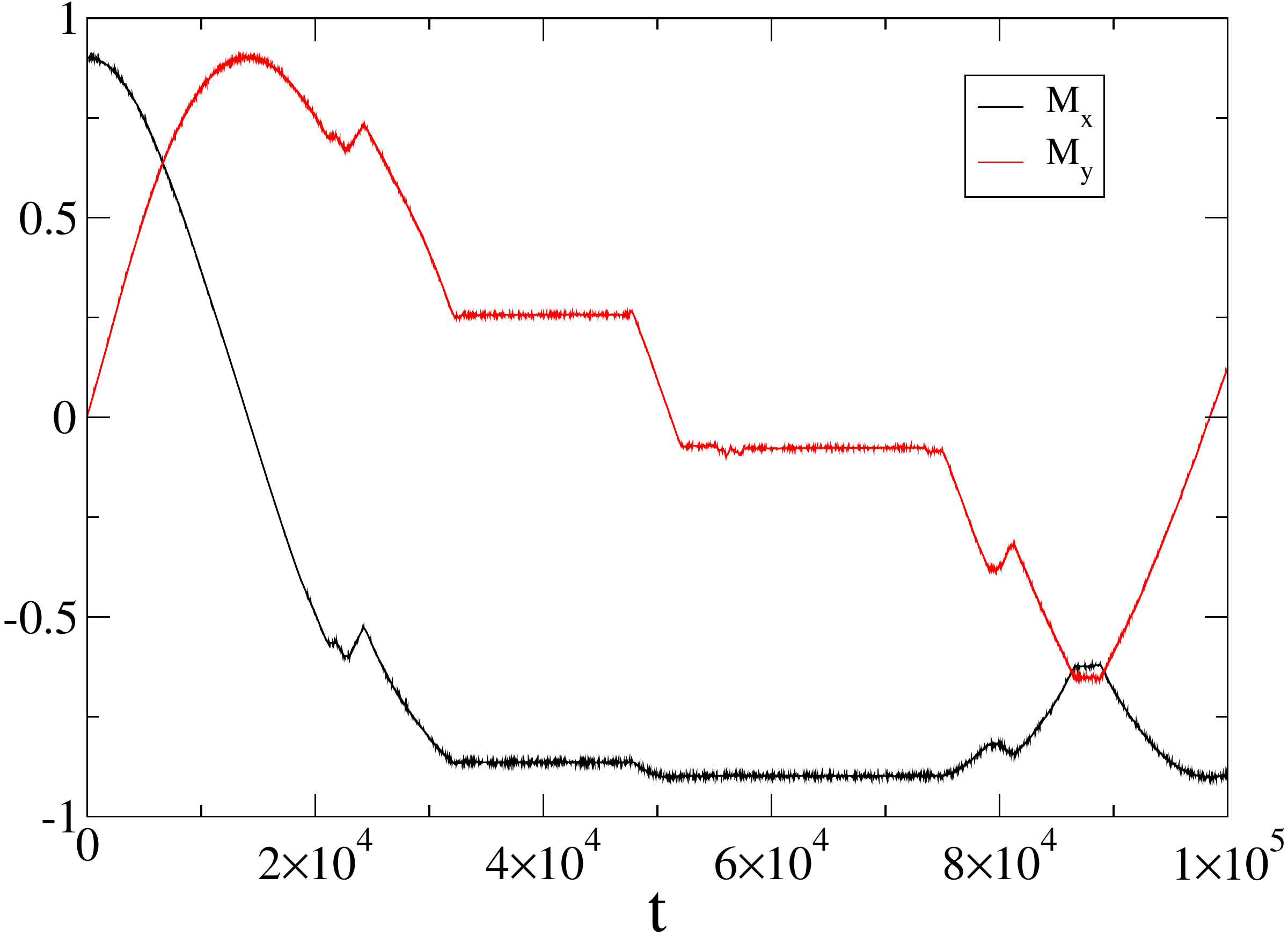}}}
\scalebox{0.3}{{\includegraphics{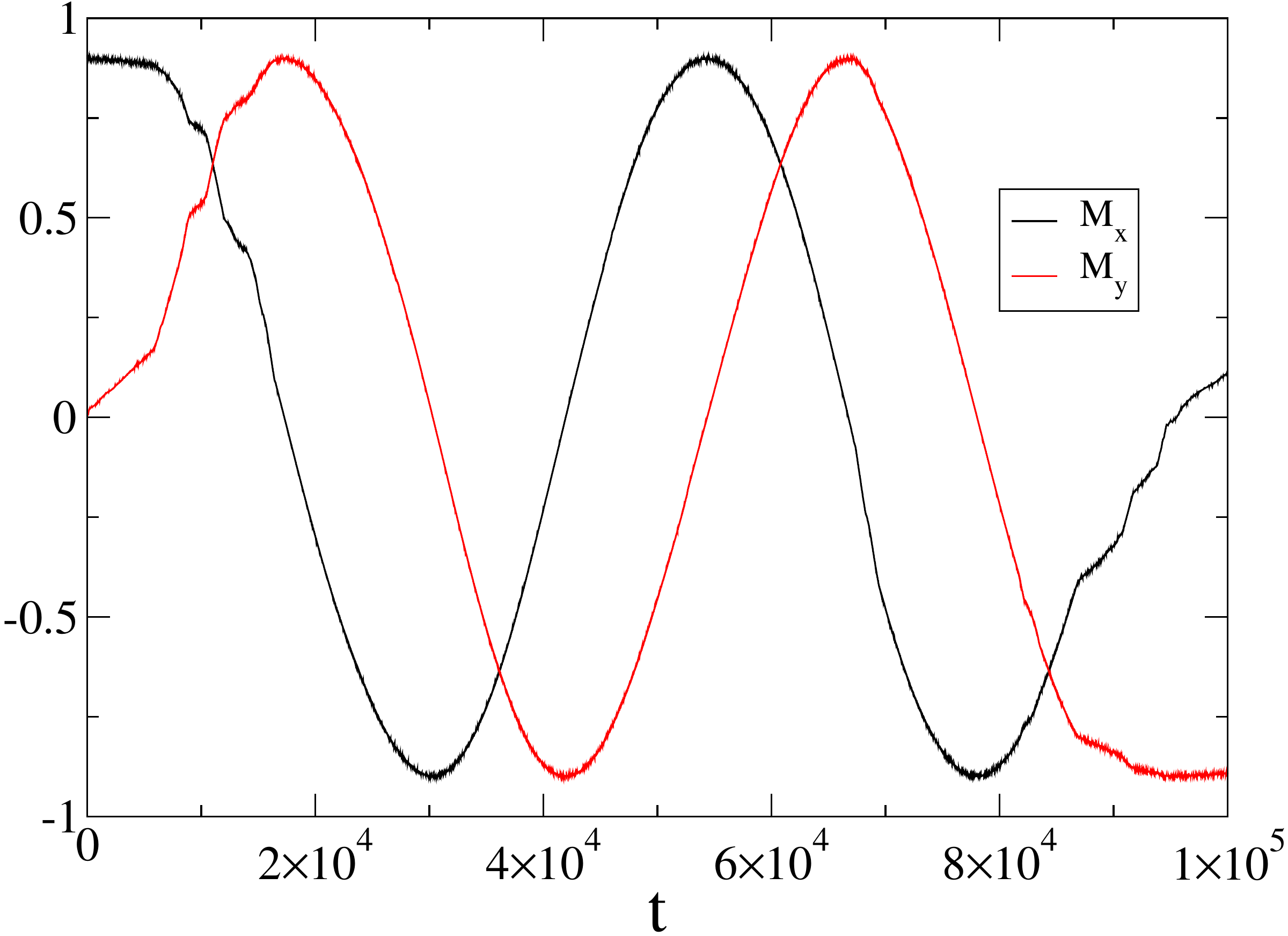}}}
\scalebox{0.3}{{\includegraphics{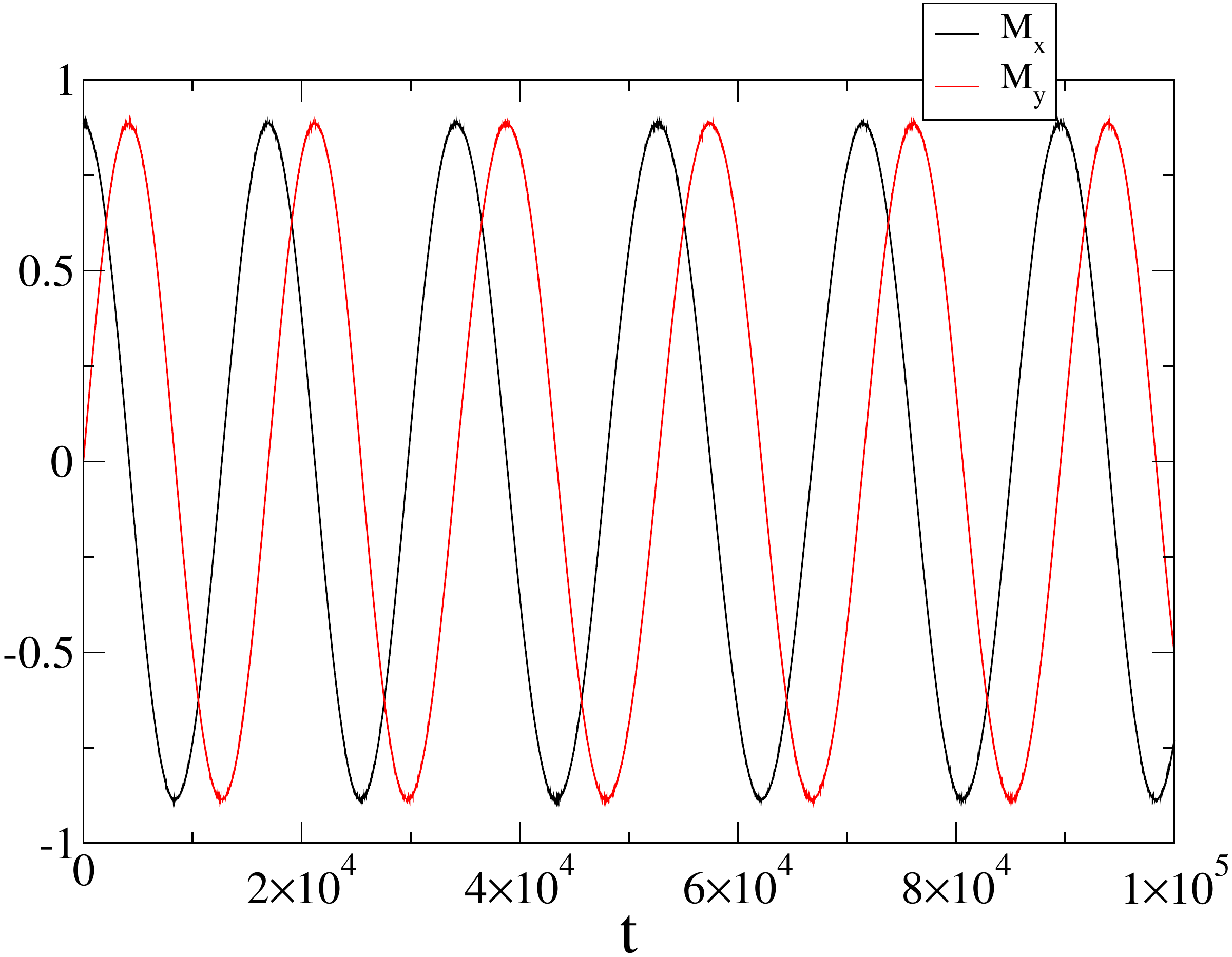}}}
\end{center}
\caption{(Color online) Magnetization components for the HMF model with $N=10\;000$ and energies per particle $e=0.17$ (a), $e=0.175$ (b),
$e=0.18$ (c) and $e=0.2$ (d).}
\label{maginstxbord}
\end{figure}

In the thermodynamic limit $N\rightarrow\infty$, the dynamics
being exactly described by a mean-field approach, the motion of each particle is statistically uncorrelated from that of
all other particles, with the force given by the mean-field force as
the statistical average of the forces due to all other particles in the system.
Let us consider the case of the HMF model where the equations of motion of particle $i$ are given by
\begin{eqnarray}
 & & \dot \theta_i=p_i,
\nonumber\\
 & & \dot p_i=-M_x\sin\theta_i+M_y\cos\theta_i=-M\sin(\theta_i+\phi).
\label{eqmotionhmf}
\end{eqnarray}
In an equilibrium or stationary state in the thermodynamic limit,
the magnetization $M$ and phase $\phi$ are constant and each particle behaves as a pendulum subject to a constant force $M$ in
the direction specified by the phase of the magnetization. As a result, all particles act as uncoupled pendula, and
the system is integrable, i.~e.\ non-chaotic. For finite $N$ the system is chaotic as its largest Lyapunov exponent~\cite{ott} does not
vanish~\cite{firpo2,lyapnos,lyapnos2}. Manos and Ruffo~\cite{manos} showed that a crossover from weak to strong chaos,
corresponding to a fraction of chaotic orbits less than $1\%$ (weak chaos) and close to $100\%$ (strong chaos), occurs at
an energy value such that the time dependence of the phase, i.~e.\ the excitation of the Goldstone mode, becomes important.
This is also reflected by the value of the Lyapunov exponent as a function of energy~\cite{manos,lyapnos,firpo2}.
In fact, for energies above the phase transition, where the magnetization vanishes in the thermodynamic limit,
the Lyapunov exponent tends to zero very fast with increasing $N$,
according to a power law $N^{-\gamma}$, with $\gamma\approx1/3$, while for energy values corresponding to strong chaos, the decrease of Lyapunov
exponent is at least one order of magnitude slower as given by the exponent $\gamma$~\cite{lyapnos}. Figure~\ref{diffcompar} at the right shows the value of the
equilibrium spatial distribution function in Eq.~(\ref{distri-space}) at $\theta=\pi$ with $\phi=0$.
If $\rho(\pi)$ is not significantly different from zero, the net flux of particles at the boundary is also very small, and
the Goldstone mode is not excited. As a consequence, no net motion of the center of mass of the system is observed
for energies below a threshold. Figure~\ref{maginstxbord} shows the behavior of the magnetization components for a few
energy values at equilibrium. A significant diffusive motion of the center of mass of the system
starts for energies greater than $e_g\approx0.17$, the energy value corresponding to the crossover from weak to strong chaos.

\begin{figure}
\begin{center}
\scalebox{0.3}{{\includegraphics{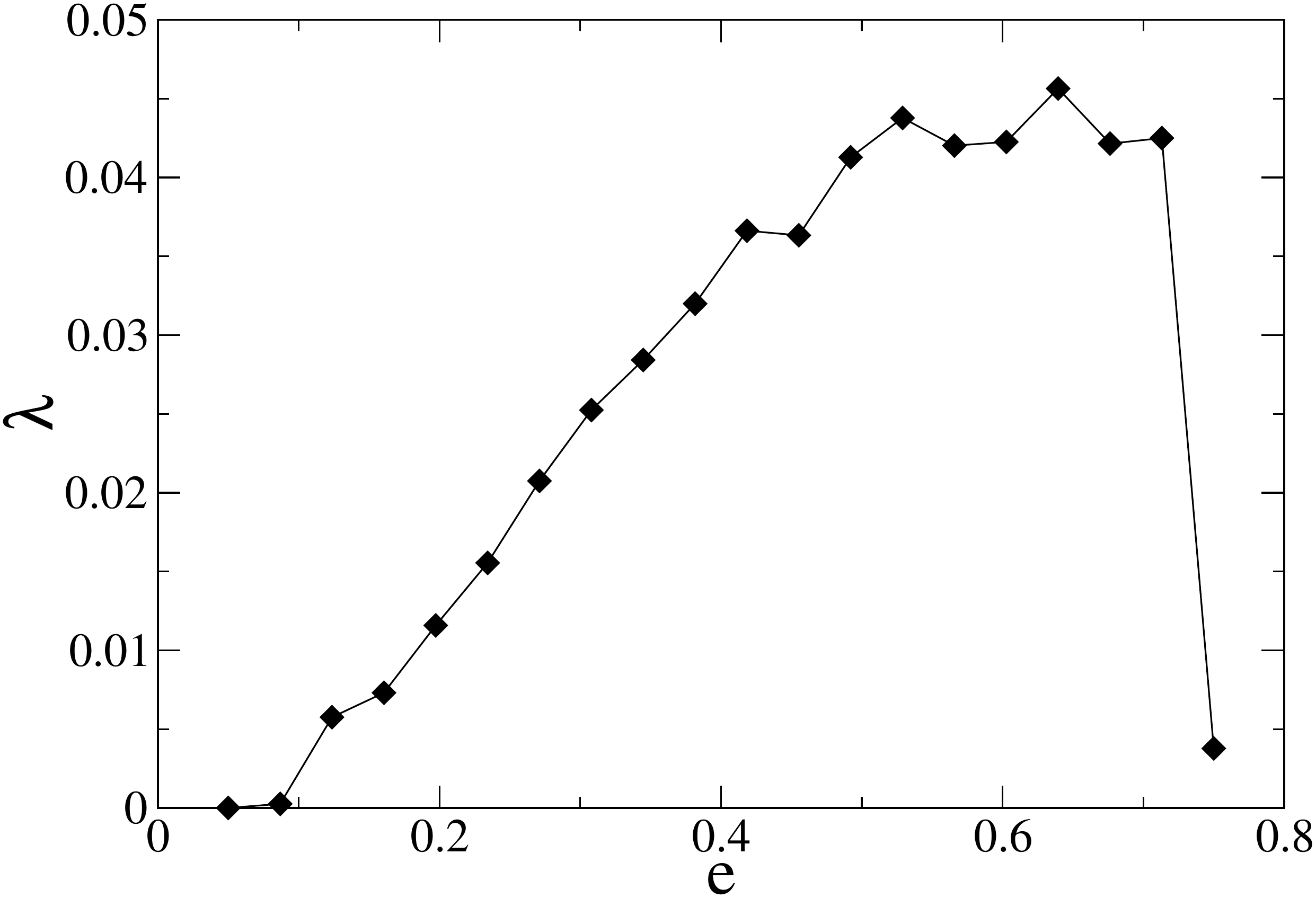}}}
\scalebox{0.3}{{\includegraphics{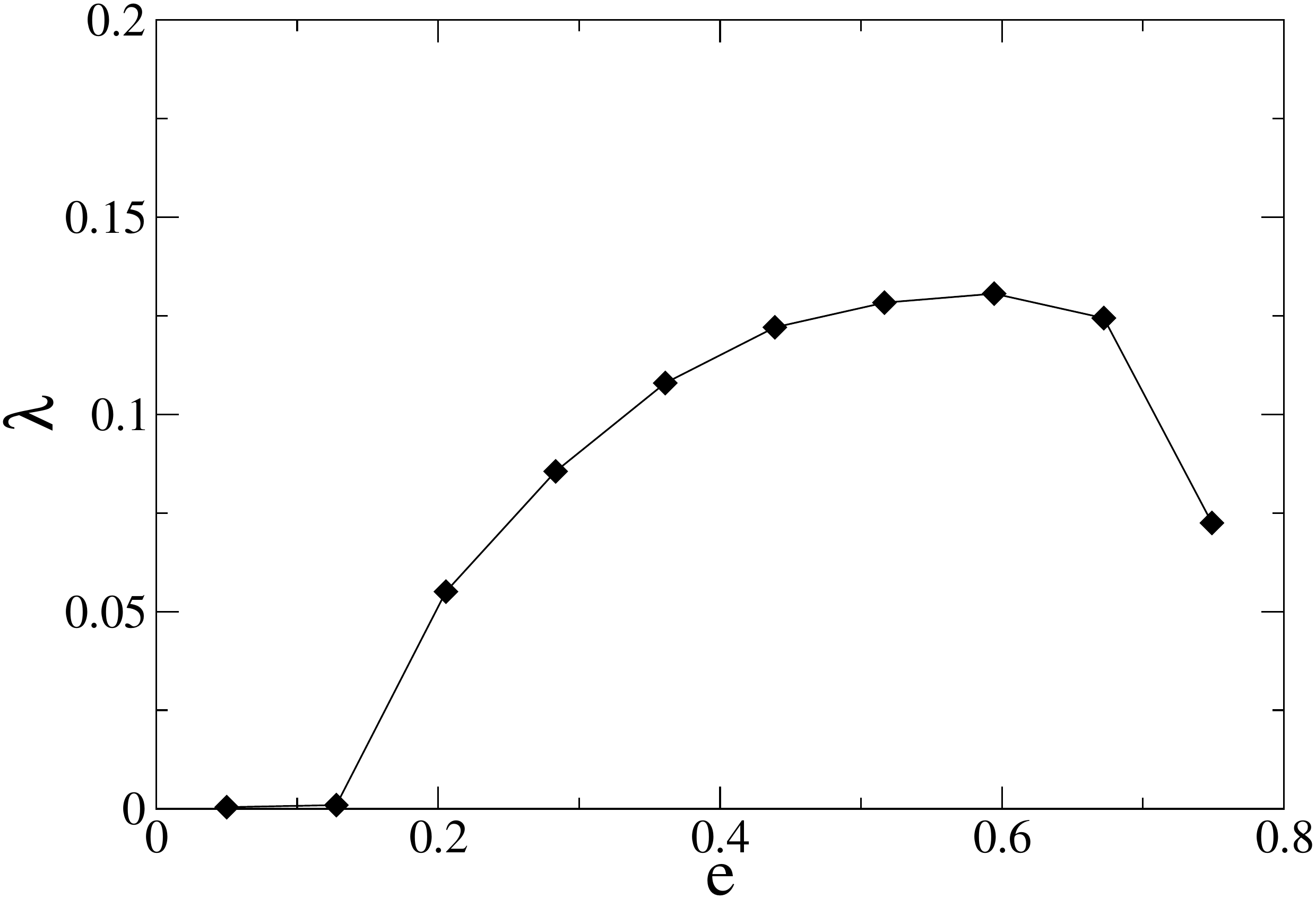}}}
\end{center}
\caption{Left panel: Largest Lyapunov exponent for the pendulum with phase given by a  Gaussian colored noise corresponding to the equation of
	motion in Eq.~(\ref{eqmotionhmf}) with $\alpha=0.01$.
	Right Panel: Largest Lyapunov exponent for the HMF model in an equilibrium state with $N=1000\,000$.}
\label{lyaposcill}
\end{figure}

In order to illustrate the relation of the coupling of the diffusive motion of the center of mass and chaos,
let us consider a single oscillator with the same equations of motion as in Eq.~(\ref{eqmotionhmf}) and phase $\phi$ given by:
\begin{equation}
\phi(K\Delta t)=\sum_{i=1}^K\Delta\phi_i,
\label{randphase}
\end{equation}
with $\Delta t$ a small fixed time interval, $K$ an integer, $\Delta\phi_i$ a realization of
an exponentially correlated colored noise, i.~e.\ given by a random variable with zero mean, a Gaussian distribution and exponential
correlation function
\begin{equation}
\langle\Delta\phi_i\Delta\phi_j\rangle=e^{-K(j-i)\alpha},
\label{randcorrfunc}
\end{equation}
with $\alpha$ constant. The variance of the Gaussian distribution of the random variable $\Delta\phi$ is chosen to be the same as the
Gaussian distribution for jumps of the center of mass of the HMF model in Eq.~(\ref{skellgauss}).
The numerical algorithm for generating such a random number is given in Ref.~\cite{fox}. The largest Lyapunov exponent can be obtained from standard
methods~\cite{parker} and is shown as a function of energy in Fig.~\ref{lyaposcill}. The dynamics of the HMF model for finite $N$ is of course much
more complex than that of a single pendulum with  constant force intensity and random phase, as different particles interact with each other and
with fluctuations in the total magnetization, creating feedback effects.
The time scales are also different, which are relevant for the magnitude of the Lyapunov exponent.
Despite that, a comparison of the graphics in Fig.~\ref{lyaposcill} with Fig.~2 of Ref.~\cite{firpo2},
shows that the coupling of the Goldstone mode to the motion of a single particle is related to the strong chaotic behavior in the non-homogeneous phase,
with the Lyapunov exponent increasing rapidly for energies above the crossover from weak to strong chaos.

It is an interesting question for further studies to understand in closer details the chaos enhancing mechanism for the HMF model
and other long-range interacting systems where the thermal excitation of a similar soft mode also occurs, such as in self-gravitating
systems and a free electron laser. This change of regime from weak to strong chaos can also be associated to the flow of particles close
to the separatrix, into and outside the region inside it, which are the particles that most contribute to the Lyapunov exponent~\cite{lyapnos2}.
This flow of particles determines the diffusive properties of the particles in the system, and therefore also that of the center of mass.

\section{Goldstone mode in other long-range systems with a periodic coordinate}
\label{secother}

We discussed above that the spontaneous symmetry breaking in a long-range interacting system leads to a Goldstone mode,
and if the spatial coordinate associated to the broken symmetry is periodic, then a diffusive motion of the center of mass
of the system ensues. To illustrate the generality of this phenomenon we show that it occurs also in two very different systems:
a self-gravitating system in two dimensions and a free electron laser.

\subsection{Two-Dimensional Self-Gravitating Systems}
\label{sec4}

\begin{figure}[ptb]
\begin{center}
\scalebox{0.28}{{\includegraphics{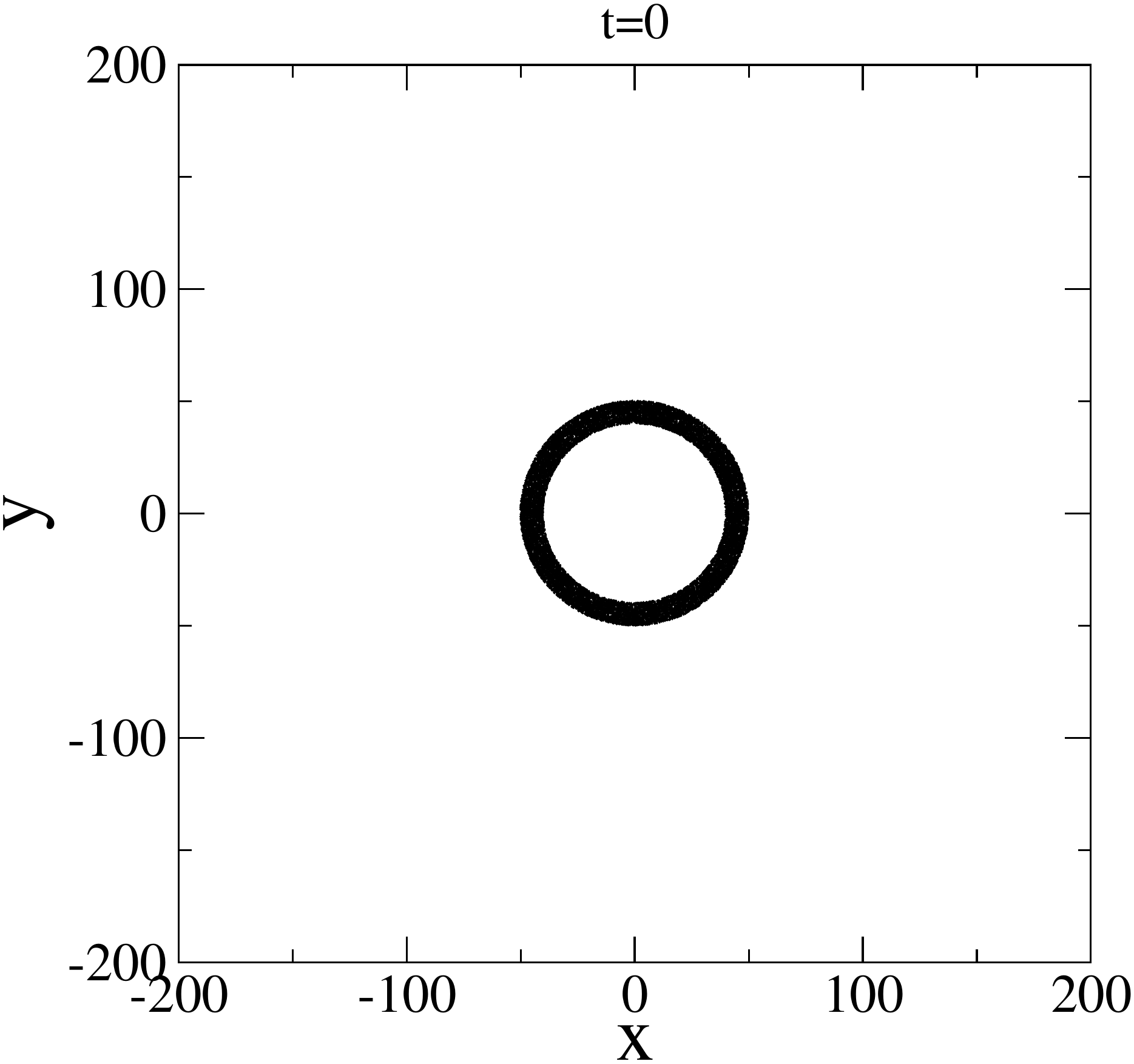}}}
\scalebox{0.28}{{\includegraphics{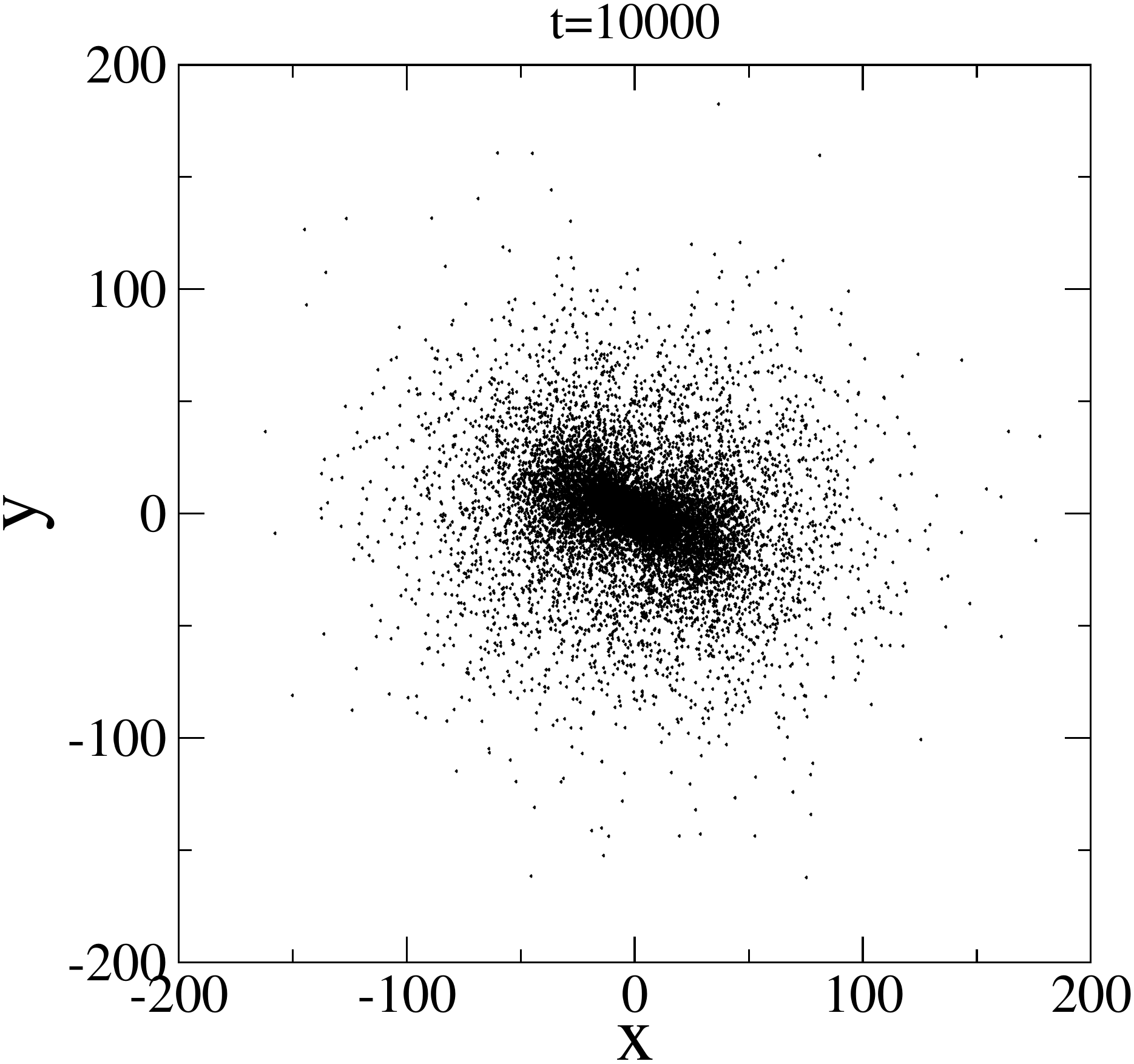}}}
\scalebox{0.28}{{\includegraphics{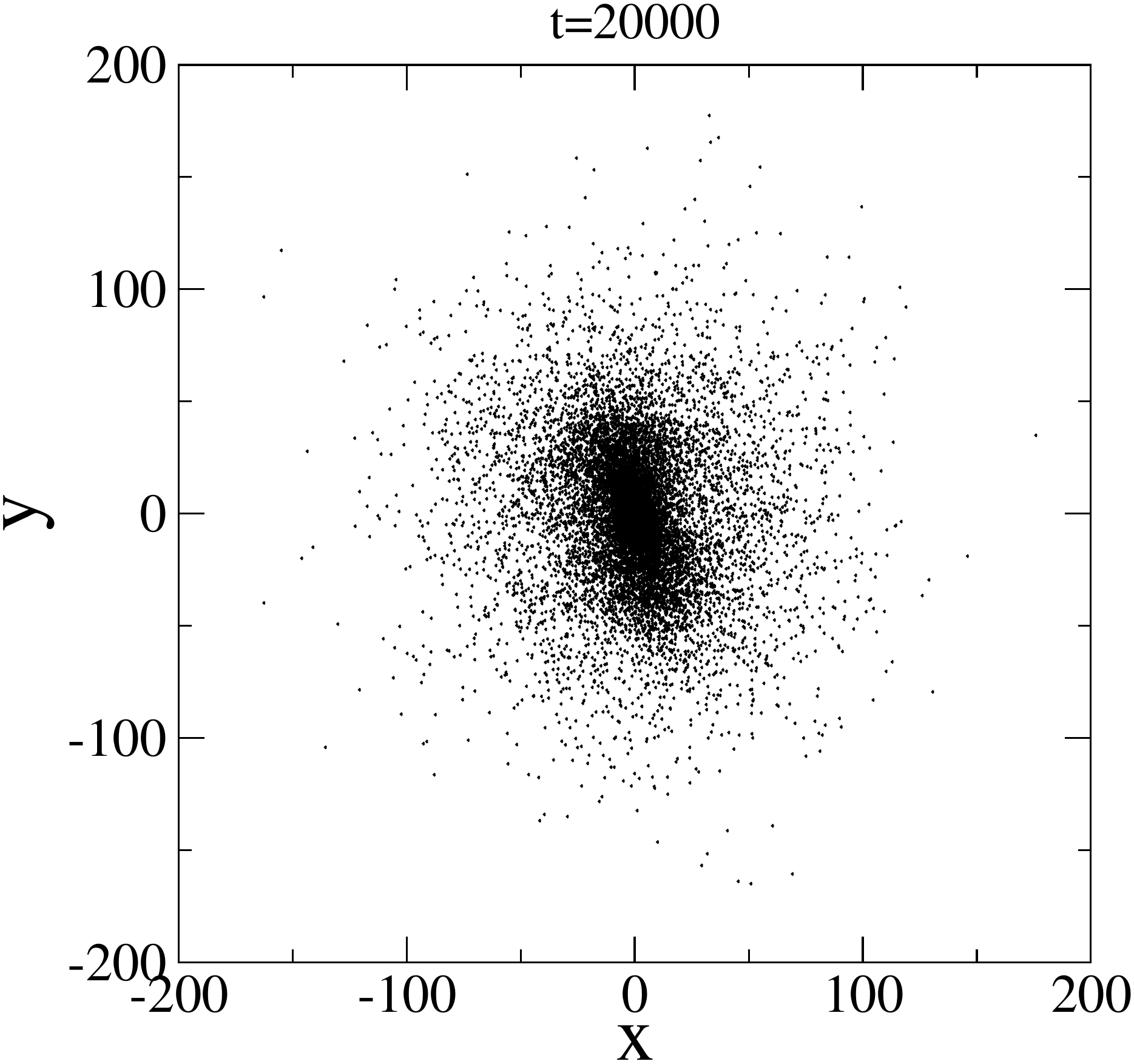}}}
\scalebox{0.28}{{\includegraphics{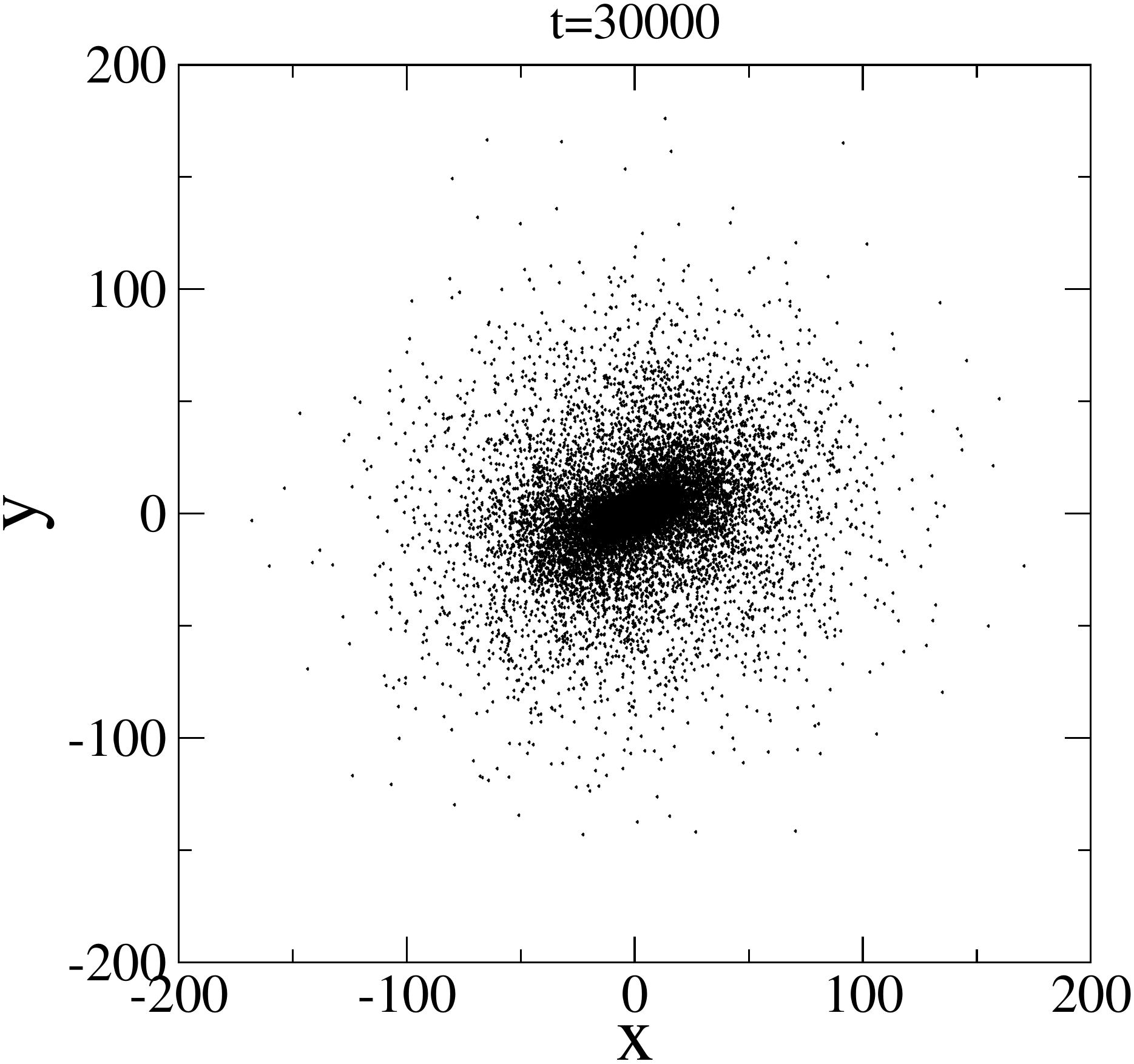}}}
\scalebox{0.28}{{\includegraphics{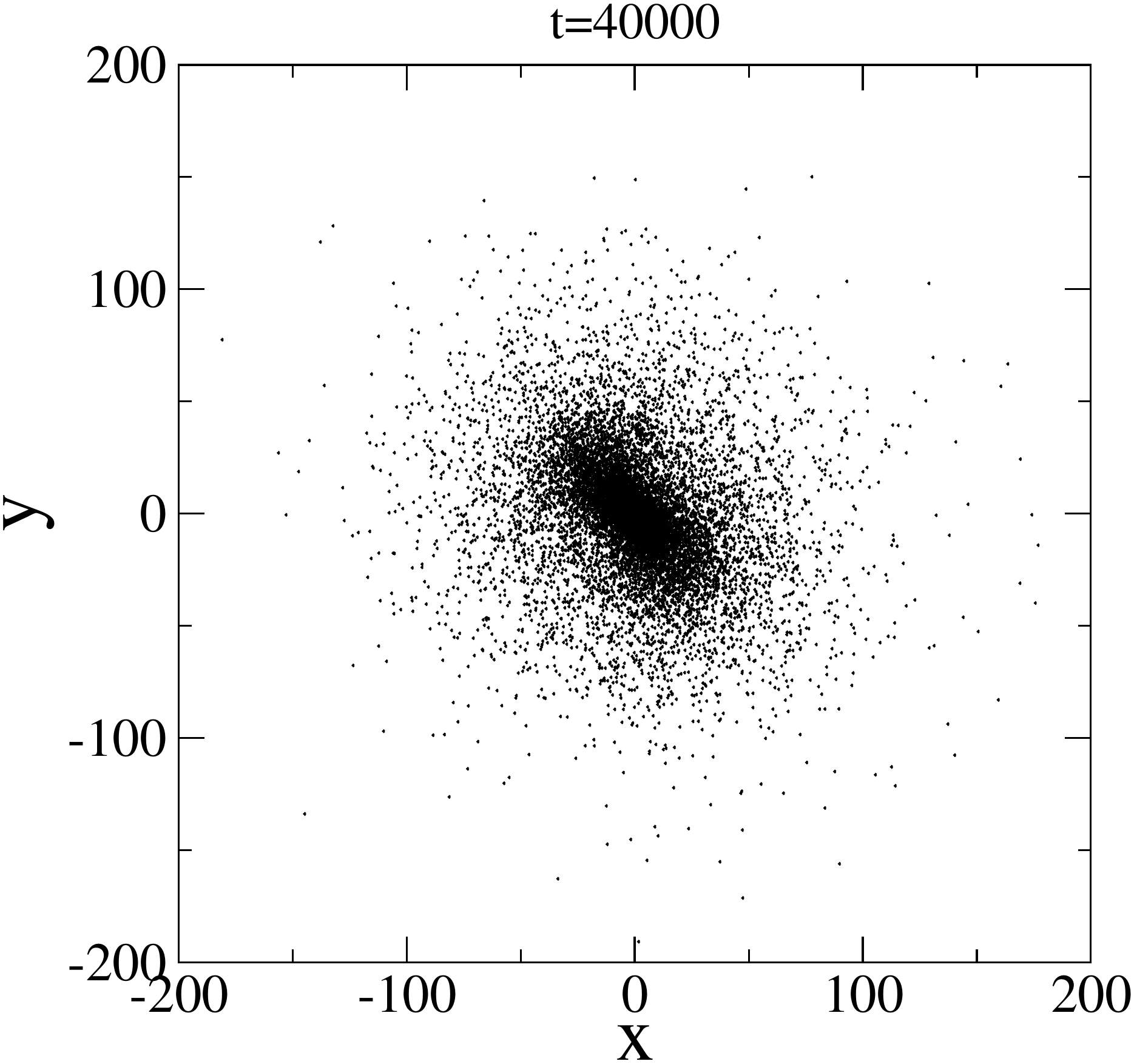}}}
\scalebox{0.28}{{\includegraphics{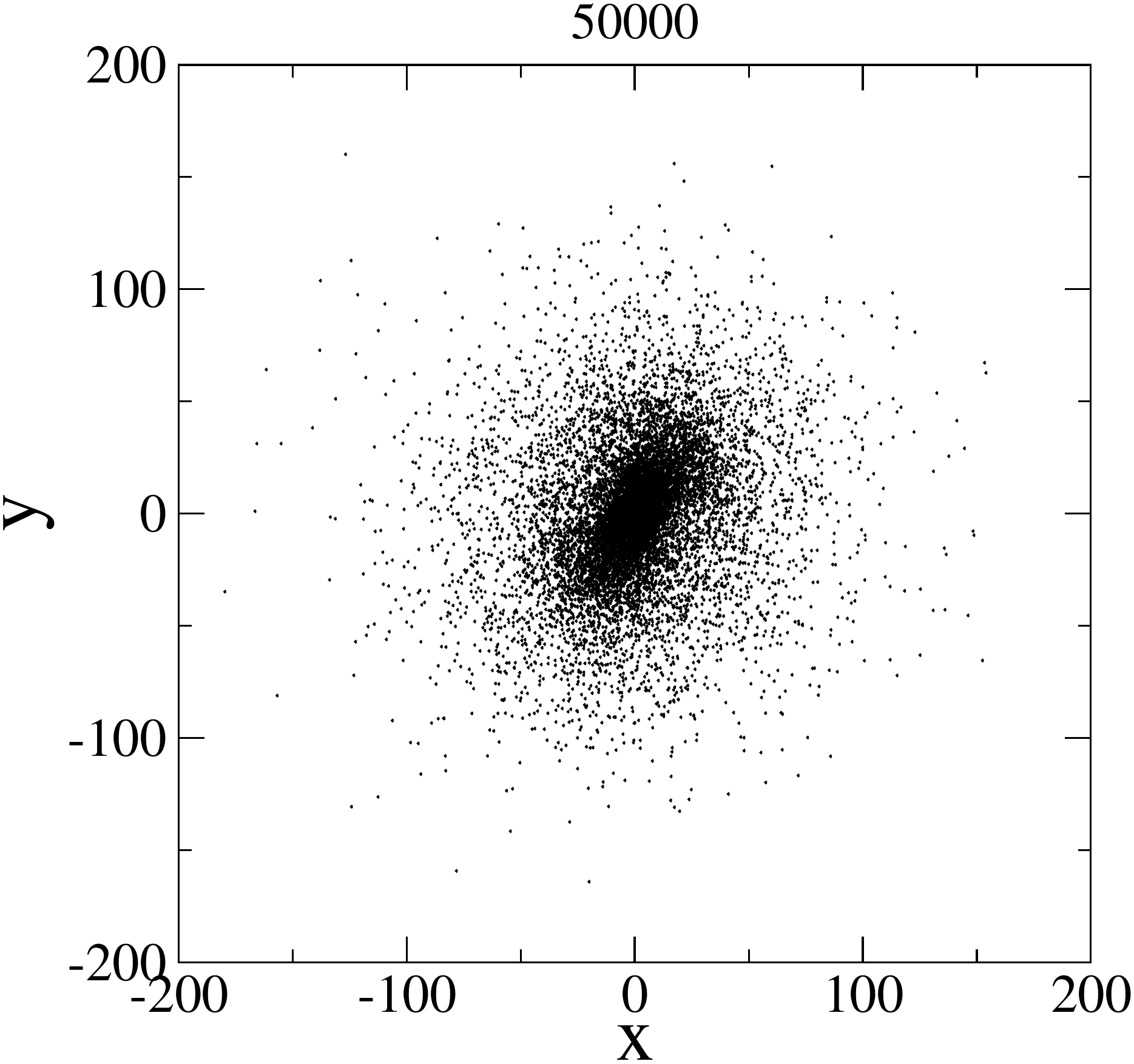}}}
\end{center}
\caption{Positions of particles in the two-dimensional self gravitating system with vanishing total angular momentum
for $N=32\:768$, time step $\Delta t=0.05$, $\epsilon=10^{-5}$ and a uniform spatial initial distribution in
a circular strip with inner and outer radius $R_1=40.0$ and $R_2=50.0$
with all particles at rest. The system evolves through the violent relaxation and reaches a
quasi-stationary state displaying a symmetry breaking.}
\label{grav2phase}
\end{figure}

\begin{figure}[ptb]
\begin{center}
\scalebox{0.3}{{\includegraphics{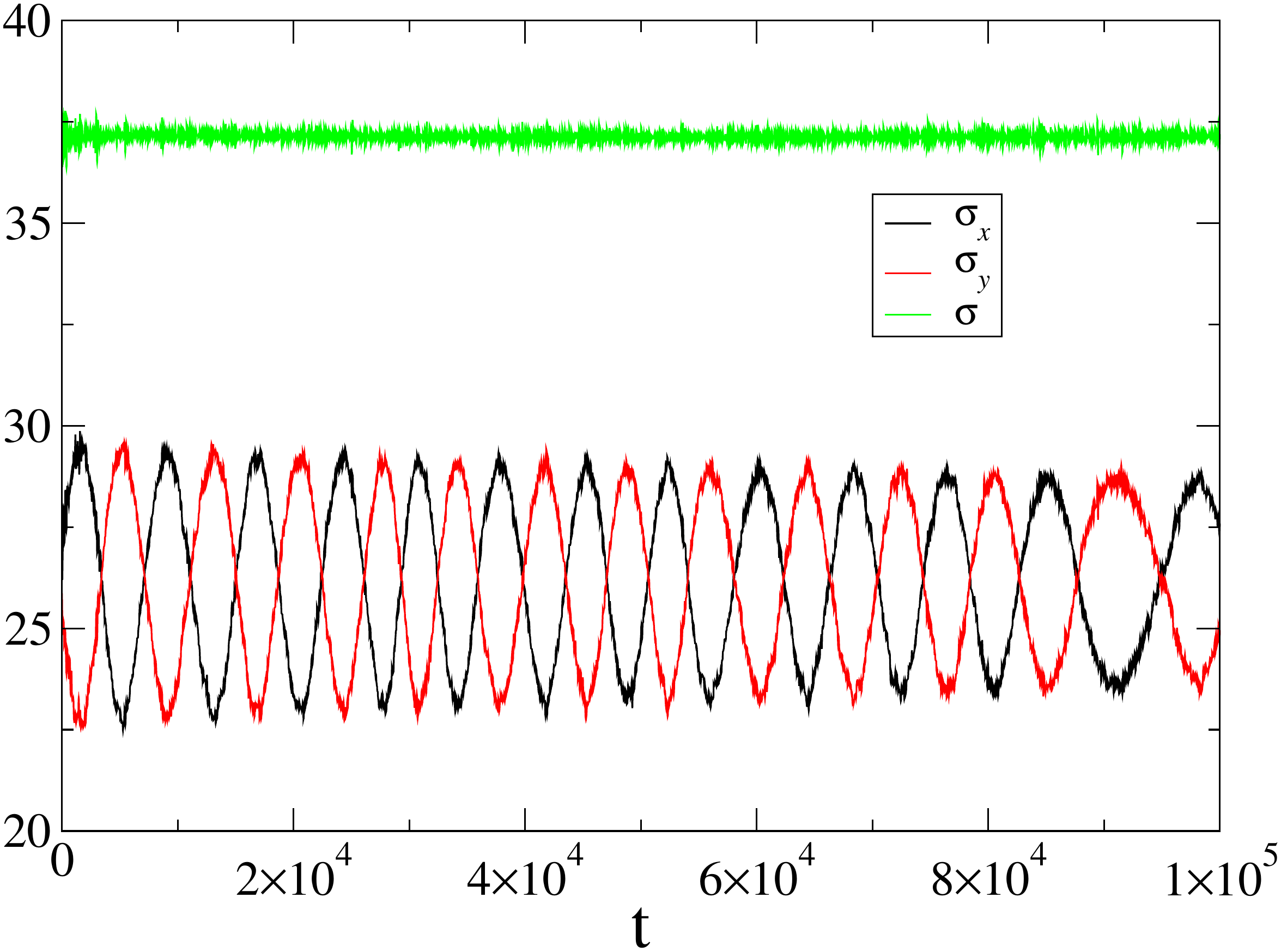}}}
\end{center}
\caption{(Color online) Position standard deviation $\sigma=\sqrt{\sigma_x^2+\sigma_y^2}$,
and standard deviations for the $x$ and $y$ coordinates for the same simulation as in Fig.~\ref{grav2phase}.
The system is initially left to evolve though the initial violent relaxation for a total time of $t=1000$.}
\label{grav2d}
\end{figure}

In order to show how generic this phenomena is we first turn our attention to two-dimensional
self-gravitating systems, with Hamiltonian~\cite{miller,telles,bruno}:
\begin{equation}
	H=\sum_{i=1}^N \frac{{\bf p}_i^2}{2}+\frac{1}{2N}\sum_{i<j=1}^N\log\left({\bf r}_i-{\bf r}_j+\epsilon\right),
\label{2dgravham}
\end{equation}
where ${\bf r}_i$ is the vector position of particle $i$ in $\mathbb{R}^2$ and ${\bf p}_i$ its conjugate momentum.
A small softening parameter $\epsilon$ was introduced in the argument of the logarithm function in Eq.~(\ref{2dgravham})
in order to avoid divergences in numerical simulations at zero inter-particle distance.
Conditions for an instability threshold for spontaneous symmetry breaking after the violent relaxation in self-gravitating systems
were discussed in~\cite{bruno1}.
We consider an initial state with all particles at rest, and spatially uniform on an annulus with inner and outer radius
$R_1$ and $R_2$, respectively. After going through a violent relaxation, the system settles on
a quasi-stationary sate with a broken rotational symmetry forming a bar structure,
as shown in Fig.~\ref{grav2phase} for some different time values, where we observe an effective (differential) rotation of the bar,
similar to what was discussed above for the HMF model.
This is caused by thermal fluctuations of the distribution function, and can be better understood by using polar coordinates
and writing down the one-particle distribution function as $f(p_r,p_\theta,r,\theta)$, where $r$ and $\theta$ are the
radial and angular coordinates, and $p_r$ and $p_\theta$ their canonically conjugate momenta, respectively.
The same reasoning as for the HMF model applies here for the angular coordinate.
The asymmetry of $f$ with respect to $\theta$ induced by momentum preserving fluctuations causes a motion of the preferred direction
with zero total angular momentum. This motion can be characterized using the inertia moments with respect to two orthogonal axis,
say $x$ and $y$, divided by the total mass, and given by:
\begin{eqnarray}
\sigma_x=\frac{1}{N}\sum_{i=1}^Nx_i^2,
\nonumber\\
\sigma_y=\frac{1}{N}\sum_{i=1}^Ny_i^2.
\label{sigdef}
\end{eqnarray}
Figure~\ref{grav2d} shows the time evolution of $\sigma_x$ and $\sigma_y$. The rotation of the system is evident albeit
the vanishing total angular momentum.

\begin{figure}[ptb]
\begin{center}
\scalebox{0.3}{{\includegraphics{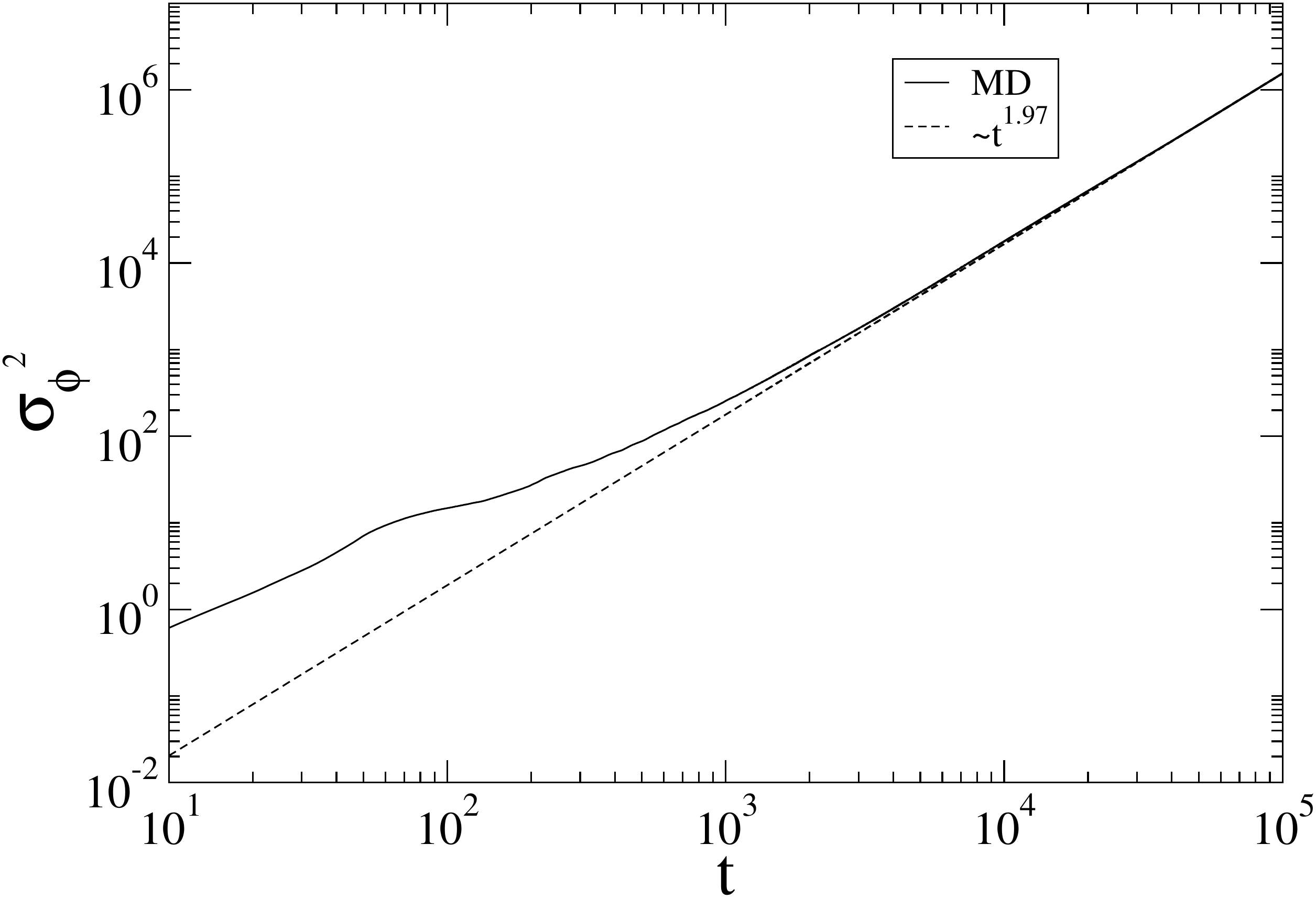}}}
\end{center}
\caption{Variance $\sigma_\phi^2$ of the angular position $\phi$ of the particles for the same simulation as in Fig.~\ref{grav2d}.
The initial position for computing the displacement $\phi(t)-\phi(0)$ is taken
at time $t=1000$, so that the initial violent relaxation has ended and the system has settled in a quasi-stationary state.
A least squares fit of a power law, shown in the figure as a dashed line, yields $\sigma^2_\theta\propto t^{1.97}$,
	i.~e.\ close to the ballistic diffusion.}
\label{grav2ddifus}
\end{figure}

This classical Goldstone mode is the outcome of a symmetry breaking with respect to a periodic coordinate, and its
motion is a result of excitations by thermal fluctuations. Since the equilibrium state has no symmetry breaking, the oscillations
for the present case are slowly damped with time and vanish once the system reaches thermodynamic equilibrium.
Figure~\ref{grav2ddifus} shows the standard deviation $\sigma_\phi$ for the position angle.
The relation in Eq.~(\ref{difcmdiffpart2}) remains valid here for the angular variable.
The position angle of the bar structure in Fig.~\ref{grav2d} varies in time with an approximately constant angular velocity, at least
for the small time window of the simulation. From the discussion in the previous section,
this is a consequence of the ballistic diffusion of the individual
particles in the angular direction. Figure~\ref{grav2ddifus} shows the variance $\sigma_\phi(t)^2=(1/N)\sum_{i=1}^N\phi_i(t)$
of the position angular variables $\phi_i(t)$, $i=1,\ldots,N$ as a function of time, 
and as expected it scales almost as $t^2$, i.~e.\ very close to ballistic diffusion.
A more detailed study of gravitational systems is beyond the scope of the present work, and will be the subject of a future publication.

\subsection{Free Electron Laser}
\label{sec5}

A Free Electron Laser is a tunable source of coherent radiation that uses a relativistic electron beam as a lasing medium.
This beam propagates in a periodic external magnetostatic field due to an undulator (or wiggler) inducing an oscillatory motion of the electrons,
which then emit synchrotron radiation that is amplified as the beam moves along the undulator~\cite{colson,bonifacio}.
Assuming a one-dimensional motion along the undulator, the equations governing the motion of the electrons in a single pass FEL
for small beam current and emittance are given by~\cite{booklri,bonifacio,yves1,yves2,yves3}:
\begin{eqnarray}
 & & \frac{d\theta_j}{dz}=p_j,
\nonumber\\
 & & \frac{dp_j}{dz}=-\sum_h F_h\left(A_he^{ih\theta_j}+A^*_he^{-ih\theta_j}\right),
\nonumber\\
 & & \frac{dA_h}{dz}=F_hb_h,
\label{eqsfel}
\end{eqnarray}
where $z$ is the distance along the undulator, $A_h=A^x_h+iA^y_h$ is the $h$-th harmonic of the field with $A^x_h$ and $A^y_h$
its transverse components, $F_h$ are coupling parameters and $b_h$ the bunching parameters given by:
\begin{equation}
b_h=-\frac{1}{N}\sum_{j=1}^N e^{-ih\theta_j}.
\label{bunching}
\end{equation}
Equations~(\ref{eqsfel}) derive from the Hamiltonian
\begin{equation}
H=\sum_{j=1}^N\frac{p_j^2}{2}-i\sum_h\sum_{j=1}^N\frac{F_h}{h}\left[A_he^{ih\theta_j}-A^*_he^{-ih\theta_j}\right],
\label{felham}
\end{equation}
with canonically conjugate variables $(\theta_j,p_j)$ and $(\sqrt{N}A_j,\sqrt{N}A^*_j)$.
The phase of the $j$-th particle with respect to the $h$-th harmonic is given by $h\theta_j$.
Here the spatial coordinate $z$ assumes the role of the time variable.
In this sense, besides the Hamiltonian in Eq.~(\ref{felham}),
the total momentum $P=\sum_jp_j+I$ is also conserved, where the total field intensity is given by $I=\sum_h|A_h|^2$. 

\begin{figure}
\begin{center}
\scalebox{0.3}{{\includegraphics{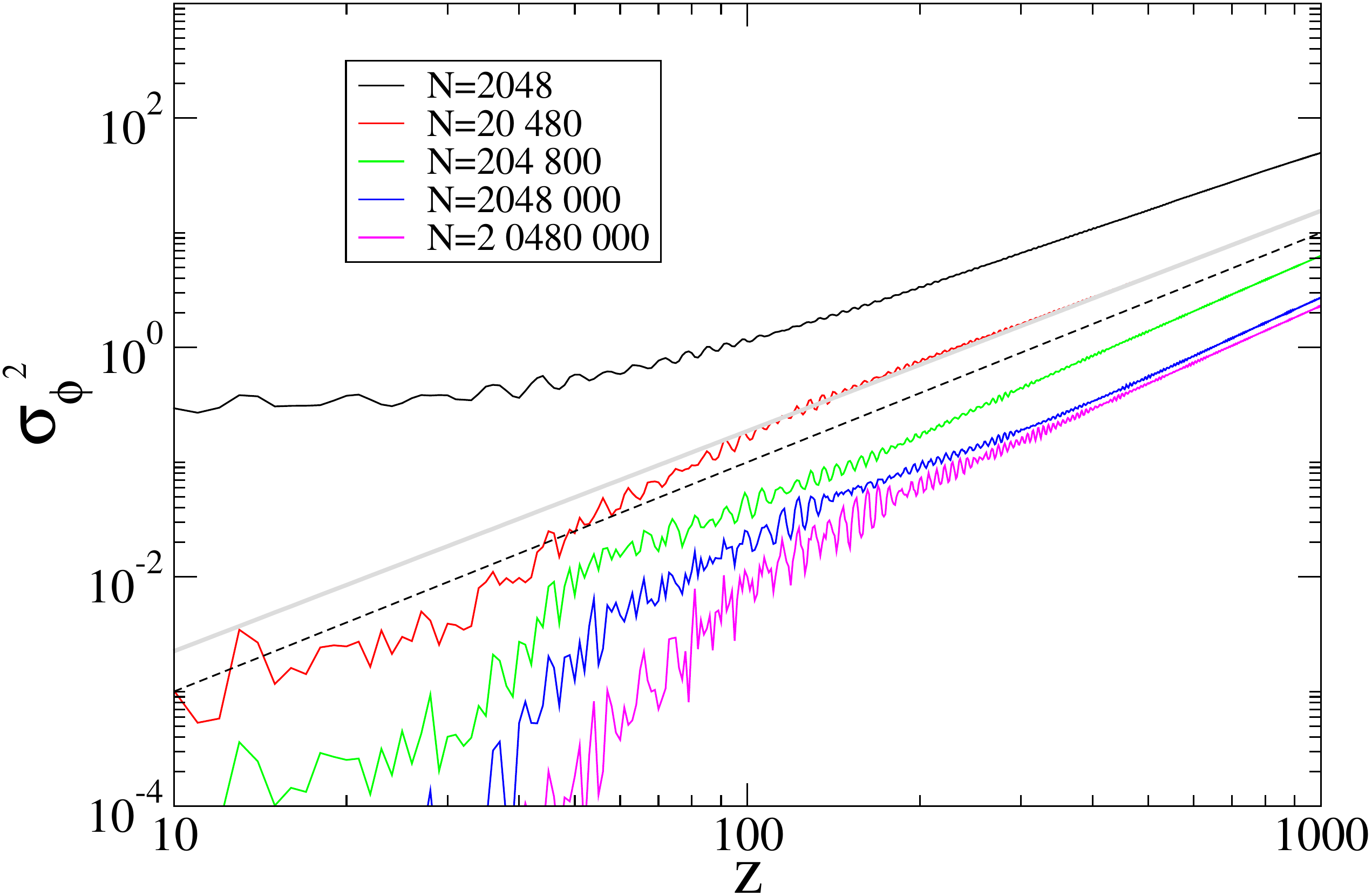}}}
\scalebox{0.3}{{\includegraphics{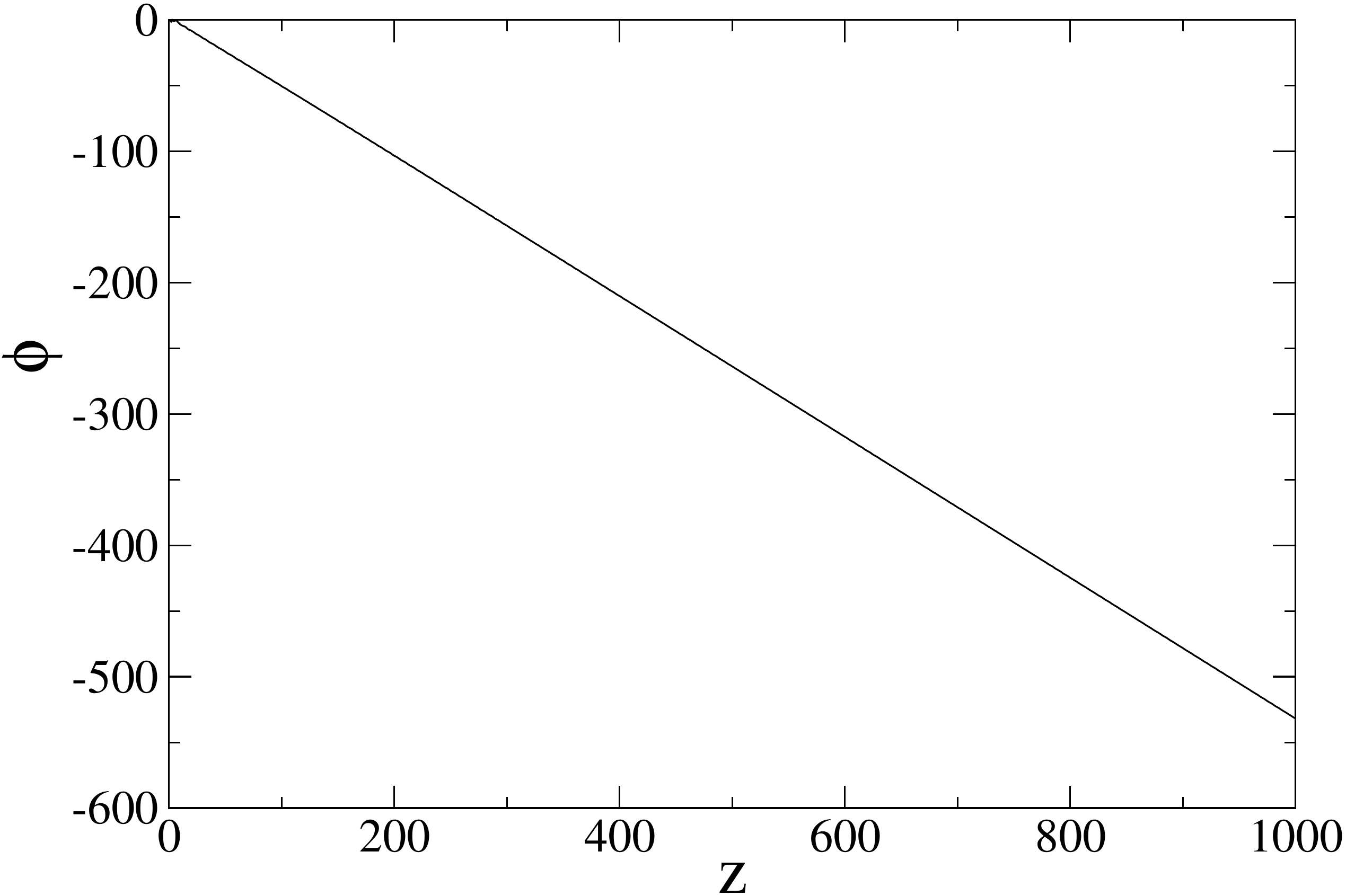}}}
\end{center}
\caption{(Color online) Left Panel: Variance $\sigma_\phi^2(z)$ in Eq.~(\ref{felvariance}) considering a single harmonic and a waterbag
initial condition with $p_0=0.5$ and $\theta_0=0.1$, for a few values of $N$ and $100$ realizations, with a time step $\Delta t=0.05$.
The dashed line introduced for reference is proportional to $z^2$.
Right Panel: value of $\phi(z)$ for one of the realizations as given by Eq.~(\ref{cmposdef}) for $N=20\,480\,000$ along the undulator.}
\label{sigfel}
\end{figure}

A diffusive motion of the center of mass of the electrons in the coordinate $\theta$ can be observed along the undulator coordinate $z$,
analogous to what we observed in the HMF model, but with non-vanishing total momentum of the electrons $\sum_jp_j$,
and approaching a constant value as the total field intensity $I$ tends to a constant. We again define the average value of the angular
coordinate using Eq.~(\ref{cmposdef}) with $z$ replacing $t$. By performing different
realizations of simulations with the same macroscopic initial conditions, the diffusion process
of the center of mass then shows up as small deviations around $\langle\phi(z)\rangle$ along the coordinate $z$, and can be quantified by the variance:
\begin{equation}
\sigma^2_\phi(z)=\left\langle\left(\phi(z)-\langle\phi(z)\rangle\right)^2\right\rangle.
\label{felvariance}
\end{equation}
The left panel of Fig.~\ref{sigfel} shows the variance $\sigma_\phi^2$ as a function of $z$,  where a superdiffusive behavior is clearly observed.
The evolution value of $\phi(z)$ for one of the realizations is shown on the right panel.

A more thorough study of this system using the methods introduced above will also be the subject of future research, as for other long-range
systems.

\section{Concluding Remarks}
\label{sec7}

We showed that, for a many-particle system with long-range interactions, if the equilibrium or a (quasi-) stationary state
spontaneously breaks a symmetry of the Hamiltonian, then a soft (Goldstone) mode exists with zero energy cost to go from one
equilibrium states to another equivalent one. Besides that, if the coordinate associated to this symmetry breaking is periodic,
this mode can be excited by thermal fluctuations due to finite $N$ effects,
resulting in a superdiffusive motion of the center of mass of the system at zero momentum,
due to the ambiguity of the position of center of mass. The existence of this soft mode was illustrated
for a two-dimensional self-gravitating system, a free electron laser and, in more details, for the HMF model. For the latter, a theory
for the ballistic motion of the center of mass was given, with expressions for relevant quantities.
An equivalent theory for more general systems rests on the development
of a theory for diffusion of non-homogeneous states, which has still to be developed. Such finite $N$ effects cannot be described
from a purely kinetic equation approach, similarly to the case of a single wave propagating in a plasma system, where separatrix crossing also
plays an important role~\cite{firpo3}.

We also discussed how the coupling of the Goldstone mode to the mean-field motion of individual particles may enhance the chaotic
behavior of the system, and illustrated this possibility again for the HMF model. This seems to be an important mechanism of chaos
enhancement in systems with long-range interactions with spontaneous symmetry breaking with respect to a periodic coordinate,
and is certainly also a point worth of further research for other similar systems.

\section{Acknowledgments}

The authors are indebted to J.B.~Fouvry for many discussions and comments. They thank S.~Ruffo for fruitful discussions and also Y.~Elskens for the long discussions and for carefully reading our manuscript.
TMRF also acknowledges partial financial support from CNPq (Brazil) grant no.\ 305842/2017-0, from Laboratoire J.A.~Dieudonn\'e and from the ``F\'ed\'eration Doeblin''. BM acknowledges support by the grant Segal ANR-19-CE31-0017 of the French Agence Nationale de la Recherche.

\end{document}